\documentclass[twocolumn]{aastex63}

\usepackage{amsmath}
\usepackage{multirow}

\received{March 20, 2020}
\revised{May 26, 2020}
\accepted{June 11, 2020}
\submitjournal{ApJ}

\shorttitle{Distinguish wet and dry TRAPPIST-1 e and f}
\shortauthors{Wunderlich et al.}
\newcommand{\pic}[2]{
	\begin{figure}
	 \plotone{#1}
	 \caption{#2}
	 \label{fig:#1}
	\end{figure}}

\newcommand{\picwide}[2]{
	\begin{figure}
	 \includegraphics[width=\columnwidth]{#1}
	 \caption{#2}
	 \label{fig:#1}
	\end{figure}}	

\newcommand{\picfull}[2]{
	\begin{figure*}
     \plotone{#1}
	 \caption{#2}
	 \label{fig:#1}
	\end{figure*}}

\newcommand{\picfullwide}[2]{
	\begin{figure*}
	 \centering
	 \includegraphics[width=0.9\textwidth]{#1}
	 \caption{#2}
	 \label{fig:#1}
	\end{figure*}}

\newcommand{\picpage}[2]{
	\begin{figure*}
	 \centering
	 \includegraphics[width=\textwidth]{#1}
	 \caption{#2}
	 \label{fig:#1}
	\end{figure*}}

\graphicspath{{./}{figures/}}

\begin{document}

\title{Distinguishing between wet and dry atmospheres of TRAPPIST-1 e and f}

\correspondingauthor{Fabian Wunderlich}
\email{fabian.wunderlich@tu-berlin.de}

\author[0000-0002-2238-5269]{Fabian Wunderlich}\thanks{Equal Contribution Authors}
\affiliation{Zentrum f\"{u}r Astronomie und Astrophysik, Technische Universit\"{a}t Berlin, 10623 Berlin, Germany}
\affiliation{Institut f\"{u}r Planetenforschung, Deutsches Zentrum f\"{u}r Luft- und Raumfahrt, 12489 Berlin, Germany}

\author[0000-0003-4331-2277]{Markus Scheucher}\thanks{Equal Contribution Authors}
\affiliation{Zentrum f\"{u}r Astronomie und Astrophysik, Technische Universit\"{a}t Berlin, 10623 Berlin, Germany}
\affiliation{Institut f\"{u}r Planetenforschung, Deutsches Zentrum f\"{u}r Luft- und Raumfahrt, 12489 Berlin, Germany}

\author[0000-0003-4770-8551]{M. Godolt}
\affiliation{Zentrum f\"{u}r Astronomie und Astrophysik, Technische Universit\"{a}t Berlin, 10623 Berlin, Germany}

\author[0000-0003-3646-5339]{J. L. Grenfell}
\affiliation{Institut f\"{u}r Planetenforschung, Deutsches Zentrum f\"{u}r Luft- und Raumfahrt, 12489 Berlin, Germany}

\author[0000-0001-7196-6599]{F. Schreier}
\affiliation{Institut f\"{u}r Methodik der Fernerkundung, Deutsches Zentrum f\"{u}r Luft- und Raumfahrt,  82234 Oberpfaffenhofen, Germany}

\author[0000-0002-5094-2245]{P. C. Schneider}
\affiliation{Hamburger Sternwarte, Gojenbergsweg 112, 21029 Hamburg, Germany}

\author[0000-0001-9667-9449]{D. J. Wilson}
\affiliation{McDonald Observatory, University of Texas at Austin, Austin, TX 78712, USA}

\author[0000-0002-0516-7956]{A. S\'{a}nchez-L\'{o}pez}
\affiliation{Sterrewacht Leiden, Universiteit Leiden, Postbus 9513, 2300 RA Leiden, The Netherlands}

\author[0000-0003-2941-7734]{M. L\'{o}pez-Puertas}
\affiliation{Instituto de Astrof\'{i}sica de Andaluc\'{i}a (IAA-CSIC), Glorieta de la Astronomía s/n, 18008 Granada, Spain}

\author{H. Rauer}
\affiliation{Zentrum f\"{u}r Astronomie und Astrophysik, Technische Universit\"{a}t Berlin, 10623 Berlin, Germany}
\affiliation{Institut f\"{u}r Planetenforschung, Deutsches Zentrum f\"{u}r Luft- und Raumfahrt, 12489 Berlin, Germany}
\affiliation{Institut f\"{u}r Geologische Wissenschaften, Freie Universit\"{a}t Berlin, 10623 Berlin, Germany}



\begin{abstract}

The nearby TRAPPIST-1 planetary system is an exciting target for characterizing the atmospheres of terrestrial planets. The planets e, f and g lie in the circumstellar habitable zone and could sustain liquid water on their surfaces. 
During the extended pre-main sequence phase of TRAPPIST-1, however, the planets may have experienced extreme water loss, leading to a desiccated mantle. The presence or absence of an ocean is challenging to determine with current and next generation telescopes. 
Therefore, we investigate whether indirect evidence of an ocean and/or a biosphere can be inferred from observations of the planetary atmosphere. 
We introduce a newly developed photochemical model for planetary atmospheres, coupled to a radiative-convective model and validate it against modern Earth, Venus and Mars. 
The coupled model is applied to the TRAPPIST-1 planets e and f, assuming different surface conditions and varying amounts of CO$_2$ in the atmosphere. As input for the model we use a constructed spectrum of TRAPPIST-1, based on near-simultaneous data from X-ray to optical wavelengths. We compute cloud-free transmission spectra of the planetary atmospheres and determine the detectability of molecular features using the Extremely Large Telescope (ELT) and the James Webb Space Telescope (JWST). 
We find that under certain conditions, the existence or non-existence of a biosphere and/or an ocean can be inferred by combining 30 transit observations with ELT and JWST within the K-band. 
A non-detection of CO could suggest the existence of an ocean, whereas significant CH$_4$ hints at the presence of a biosphere.

\end{abstract}



\section{Introduction} \label{sec:intro}
The nearby TRAPPIST-1 system offers exciting new opportunities for studying the atmospheres of its seven planets with next generation telescopes such as the JWST \citep[James Webb Space Telescope; ][]{gardner2006,beichman2014} or the ELT \citep[European Large Telescope; ][]{gilmozzi2007}.
Due to short orbital periods and large star-planet contrast ratios, planets orbiting such cool host stars are easier to detect and characterize via the transit method than planets orbiting hotter stars and are therefore prime targets to observe the properties of their atmospheres.  

On the other hand the stellar luminosity evolution of M-dwarfs is quite different to that of solar-type stars. In particular the active pre-main sequence phase of the star can be extended and the stellar Ultra Violet (UV) radiation is high for about a billion years \citep[see e.g.][]{baraffe2015,luger2015}. This could lead to a runaway greenhouse state on an ocean-bearing terrestrial planet and a loss of substantial amounts of planetary water vapour (H$_2$O) before the star enters the main sequence phase \citep[see e.g.][]{wordsworth2013, ramirez2014, luger2015,tian2015, bolmont2017,bourrier2017}. 
Recently \citet{fleming2020} suggest, that TRAPPIST-1 has maintained high activity with a saturated XUV luminosity (X-ray and extreme UV emission) for several Gyrs. Hence, the planets likely received a persistent and strong XUV flux from the host star for most of their lifetimes.

In such an environment with strong H$_2$O photolysis and subsequent hydrogen escape it has been suggested that the atmosphere could build up thousands of bar molecular oxygen (O$_2$) when assuming e.g. inefficient atmospheric loss or surface sinks \citep{wordsworth2014, luger2015,lincowski2018}. This build-up can be prevented if O$_2$ is absorbed into the surface during the early magma ocean phase \citep[see e.g.][]{schaefer2016, wordsworth2018} or by extreme UV driven oxygen escape \citep{tian2015b,dong2018,guo2019,johnstone2020}. \citet{grenfell2018} suggest, that if enough molecular hydrogen (H$_2$) is present it can react with O$_2$ from H$_2$O photolysis to reform water via explosion-combustion reactions.


\citet{bolmont2017} concluded that the TRAPPIST-1 planets can retain significant amount of water even for strong far UV (FUV) photolysis of H$_2$O and large hydrogen escape rates. Three (TRAPPIST-1 e, f, and g) of the seven planets lie in the classical habitable zone (HZ), defined as the region around the star where a planet could maintain liquid water on its surface \citep{kasting1993}. 3D simulations show that only TRAPPIST-1~e would allow for surface liquid water without the need of greenhouse warming from a gas other than H$_2$O \citep{wolf2017,turbet2018}. The other two planets require greenhouse gases such as carbon dioxide (CO$_2$) and thick atmospheres to sustain surface habitability \citep{turbet2018}. 

The large FUV to near UV (NUV) stellar flux ratio of TRAPPIST-1 favors abiotic build-up of O$_2$ and O$_3$ in CO$_2$-rich atmospheres \citep[e.g.][]{tian2014}. Hence, O$_2$ or ozone (O$_3$) cannot be considered as reliable biosignature gases like on Earth \citep[e.g.][]{selsis2002, segura2007, harman2015, meadows2017}. Due to weak stellar UV emissions at wavelengths longer than 200~nm, planets orbiting M-stars show an increase in the abundance of certain bioindicators and biomarkers such as methane (CH$_4$) and nitrous oxide (N$_2$O) compared to the Earth around the Sun \citep[see][]{segura2005, rauer2011, grenfell2013, rugheimer2015, wunderlich2019}. Assuming the same surface emissions as on Earth, CH$_4$ would be detectable with the JWST in the atmosphere of a habitable zone Earth-like planet around TRAPPIST-1 \citep{wunderlich2019}. \citet{krissansen2018a} argued that the simultaneous detection of CH$_4$ and CO$_2$ in the atmosphere of a planet in the HZ is a potential biosignature. However, the build-up of detectable amounts of CH$_4$ is also conceivable by large outgassing from a more reducing mantle than Earth.

The detection of CO$_2$ in cloud-free atmospheres of TRAPPIST-1 planets would be feasible within approximately ten transits with the JWST  \citep[see][]{morley2017, krissansen2018, wunderlich2019, lustig-yaeger2019, fauchez2019,komacek2020}. The detection of other species, such as O$_3$ would require many more transits \citep[see e.g.][]{lustig-yaeger2019, fauchez2019,pidhorodetska2020}. Another species which might be detectable in CO$_2$-rich atmospheres is carbon monoxide (CO), produced by CO$_2$ photolysis \citep[e.g.][]{schwieterman2019}. Since CO has only a few abiotic sinks and weak biogenic sources it is often considered as a potential antibiosignature \citep{zahnle2008, wang2016, nava2016, meadows2017, catling2018}.

\citet{wang2016} argued that simultaneous observations of O$_2$ and CO would distinguish a true biosignature (O$_2$ without CO) from  a photochemically  produced  false  positive  biosignature (O$_2$ with CO). However, \citet{rodler2014} showed that a detection of Earth-like O$_2$ levels with ELT would only be feasible for a planet around a late M-dwarf at a distance below $\sim$5~pc \citep[see also][]{snellen2013,brogi2019,serindag2019}. 

In this study we investigate how the presence of an ocean as an efficient sink for CO would affect the atmospheric concentration of CO and other species. We simulate transmission spectra of TRAPPIST-1~e and TRAPPIST-1~f and determine the detectability of molecular features with the upcoming space-borne telescope JWST and the next generation ground-based telescope ELT. For the JWST we consider low resolution spectroscopy (LRS) and for the ELT we use high resolution spectroscopy (HRS). 
In particular we show how much CO$_2$ would be needed to obtain a detectable CO feature in a desiccated atmosphere of TRAPPIST-1~e. 

Also the photochemical processes related to the existence of a water reservoir may change the abundances of CO and O$_2$. The recombination of CO and atomic oxygen into CO$_2$ via catalytical cycles was suggested to be slower for dry CO$_2$ atmospheres due to the lower abundances of hydrogen oxides, HO$_x$ (defined as H + OH + HO$_2$) \citep[see e.g.][]{selsis2002, segura2007, krissansen2018a}.

We use a 1D climate-photochemistry model to calculate the composition profiles of CO and other species such as O$_2$ and O$_3$ in CO$_2$-poor and CO$_2$-rich atmospheres. In order to consistently simulate the photochemical processes in CO$_2$-dominated atmospheres we introduced extensive model updates. The stellar Spectral Energy Distribution (SED) is an input for the model. The UV range of the SED is crucial for the photochemical processes in the atmosphere. To our knowledge we are the first study using an SED of TRAPPIST-1 constructed based on measurements in the UV \citep{wilson2020}. For comparison we also investigate two other SEDs of TRAPPIST-1 with modelled or estimated UV fluxes as input for our climate-photochemistry model. 

In Section~\ref{sec:model} we introduce the climate-photochemistry model and validate the new version by calculating the atmospheres of modern Earth, Venus and Mars. We compare the results with other photochemical models and available observations. We also describe the line-by-line spectral model used to simulate transmission spectra of TRAPPIST-1~e and TRAPPIST-1~f, and introduce the calculation of the signal to noise ratio (S/N) of atmospheric molecular features. In Section ~\ref{ssec:sed_scenarios} we show the TRAPPIST-1 SEDs used in this study and the considered atmospheric scenarios. Results of the atmospheric modelling, simulated transmission spectra and S/N calculations are presented in Section~\ref{sec:results}. In Section~\ref{sec:discussion} we discuss our results and in Section~\ref{sec:conclusion} we present the summary and conclusion.

\section{Methodology} \label{sec:model}

\subsection{Climate-chemistry model} 
To simulate the potential atmospheric conditions of the habitable zone planets TRAPPIST-1~e and TRAPPIST-1~f we use a 1D steady-state, cloud-free, radiative-convective photochemical model, entitled 1D-TERRA. The code is based on the model of \citet{kasting1986, pavlov2000, segura2003} and was further developed by e.g. \citet{vonparis2008, vonparis2010, rauer2011,vonparis2015,gebauer2018}. We have extensively modified both the radiative-convective part of the model as well as the photochemistry module. The updated version of the model is capable of simulating a wide range of atmospheric temperatures (100~-~1000~K) and pressures (0.01~Pa - 10$^3$~bar). It covers a wide range of atmospheric compositions including potential habitable terrestrial planets, having N$_2$, CO$_2$, H$_2$ or H$_2$O-dominated atmospheres. The climate module is briefly described in Section~\ref{ssec:climate_model}. For a detailed description of the climate module we refer to the companion paper by \citet{scheucher2020}. Here we give a detailed description of the updated photochemistry model in Section \ref{ssec:photo_model}. 

\subsection{Climate module} \label{ssec:climate_model}
The atmospheric temperature for each of the pressure layers is calculated with our climate module. The radiative transfer module REDFOX uses a flexible k-distribution model for opacity calculations based on the random-overlap assumption \citep[see][]{scheucher2020}. The radiative transfer is solved using the two-stream approximation \citep{toon1989}. The module considers 20 absorbers from HITRAN 2016   \citep{gordon2017} as well as 81 absorbers in the visible (VIS) and ultraviolet (UV) with cross sections taken from the MPI Mainz Spectral Atlas \citep{keller2013}, the JPL Publication No. 15-10 \citep{burkholder2015}, \citet{mills1998} and \citet{zahnle2008}. 

Additionally, REDFOX includes Collision-Induced Absorption (CIA) data from HITRAN\footnote{www.hitran.org/cia/ \citep{karman2019}} and MT\_CKD continua from \citet{mlawer2012}.
Rayleigh scattering is considered using calculated cross sections of CO, CO$_2$, H$_2$O, N$_2$ and O$_2$ \citep{allen1973} and measured cross sections of He, H$_2$ and CH$_4$ \citep{shardanand1977}. 

To calculate the H$_2$O profile up to the cold trap we either use the relative humidity profile of the Earth taken from \citet{manabe1967} or we use a constant relative humidity throughout the troposphere. Above the cold trap the H$_2$O profile is calculated with the chemistry module. \citet{godolt2016} showed that for surface temperatures warmer than the mean surface temperature of the Earth, the relative humidity profile of \citet{manabe1967} underestimates H$_2$O abundances in the troposphere compared to 3D studies, hence, the warming due to H$_2$O absorption would also be underestimated. 

\subsection{Photochemistry module BLACKWOLF} \label{ssec:photo_model}

\begin{deluxetable}{p{1.5cm} p{6cm}} \label{tab:species}
    \tabletypesize{\footnotesize}
    \tablecolumns{2}
    \tablecaption{Species considered in the photochemical model.}
    \tablehead{\colhead{Atoms} & \colhead{Species}}  
    \startdata
O, H & O, O($^1$D), O$_2$, O$_3$, H, H$_2$, OH, H$_2$O, HO$_2$, H$_2$O$_2$ \\
C, H & C, C$_2$, CH, CH$^3_2$, CH$^1_2$, CH$_3$, CH$_4$, C$_2$H, C$_2$H$_2$, C$_2$H$_3$, C$_2$H$_4$, C$_2$H$_5$, C$_2$H$_6$, C$_3$H$_2$, C$_3$H$_3$, CH$_2$CCH$_2$, CH$_3$C$_2$H, C$_3$H$_5$, C$_3$H$_6$, C$_3$H$_7$, C$_3$H$_8$, C$_4$H, C$_4$H$_2$, C$_5$H$_4$ \\ 
C, O, H & CO, CO$_2$, HCO, H$_2$CO, H$_3$CO, CH$_3$OH, HCOO, HCOOH, CH$_3$O$_2$, CH$_3$OOH, C$_2$HO, C$_2$H$_2$O, CH$_3$CO, C$_2$H$_3$O, CH$_3$CHO, C$_2$H$_5$O, C$_2$H$_5$CHO \\
N, O & N, N$_2$, NO, NO$_2$, NO$_3$, N$_2$O, N$_2$O$_5$ \\
N, O ,H, C & NH, NH$_2$, NH$_3$, HNO, HNO$_2$, HNO$_3$, HO$_2$NO$_2$, CN, HCN, CNO, HCNO, CH$_3$ONO, CH$_3$ONO$_2$, CH$_3$NH$_2$, C$_2$H$_2$N, C$_2$H$_4$NH, N$_2$H$_2$, N$_2$H$_3$, N$_2$H$_4$ \\
S, O & S, S$_2$, S$_3$, S$_4$, S$_5$, S$_6$, S$_7$, S$_8$, SO, SO$_2$, SO$^1_2$, SO$^3_2$, SO$_3$, S$_2$O, S$_2$O$_2$ \\
S, O, H, C & HS, H$_2$S, HSO, HSO$_2$, HSO$_3$, H$_2$SO$_4$, CS, CS$_2$, HCS, CH$_3$S, CH$_4$S, OCS, OCS$_2$ \\ 
Cl, O & Cl, Cl$_2$, ClO, OClO, ClOO, Cl$_2$O, Cl$_2$O$_2$ \\
Cl, O, H, N, S & HCl, CH$_2$Cl, CH$_3$Cl, HOCl, NOCl, ClONO, ClONO$_2$, COCl, COCl$_2$, ClCO$_3$, SCl, ClS$_2$, SCl$_2$, Cl$_2$S$_2$, OSCl, ClSO$_2$
    \enddata
    \tablecomments{Each specie only appears once.}
\end{deluxetable}

\begin{deluxetable*}{lllcc} \label{tab:reactions}
    \tabletypesize{\footnotesize}
    \tablecolumns{8}
    \tablecaption{Reaction rates of bi-molecular reactions (R) in cm$^3$ s$^{-1}$, termolecular reactions (M) in cm$^6$ s$^{-1}$, thermo-dissociation reactions (T) in s$^{-1}$, and quantum yields of photolysis reactions (P) used in the photochemical module. }
    \tablehead{No. & \colhead{Reaction}  & \colhead{Reaction rate or quantum yield} & \colhead{Temperature} & \colhead{Reference}}  
    \startdata
    R1 & C + H$_2$S $\rightarrow$ CH + HS & $2.1 \cdot 10^{-10}$ & 298 & NIST \\
    R2 & C + O$_2$ $\rightarrow$ CO + O & $5.1 \cdot 10^{-11} \cdot (T/298.0)^{-0.3}$ & 15 - 295 & NIST \\
    R3 & C + OCS $\rightarrow$ CO + CS & $1.01 \cdot 10^{-10}$ & 298 & NIST \\
    \hline
    M1 & C + H$_2$ + M $\rightarrow$ CH$^3_2$ + M & $k_0 = 7.0 \cdot 10^{-32}$ & 300 & \citet{moses2011} \\
     & &  $k_{\infty} = 2.06 \cdot 10^{-11} \cdot e^{-57.0/T}$ &  &  \\
    M2 & CH$_3$ + CH$_3$ + M $\rightarrow$ C$_2$H$_6$ + M & $k_0 = 1.68 \cdot 10^{-24} \cdot (T/298.0)^{-7.0} \cdot e^{-1390.0/T}$ & 300 - 2000 & \citet{sander2011} \\
     & &  $k_{\infty} = 6.488 \cdot 10^{-11} \cdot (T/298.0)^{-0.5} \cdot e^{-25.0/T}$ & &  \\
    M3 & CH$_3$ + O$_2$ + M $\rightarrow$ CH$_3$O$_2$ + M & $k_0 = 4.0 \cdot 10^{-31} \cdot (T/298.0)^{-3.6}$ & 200 - 300 & NIST \\
     & &  $k_{\infty} = 1.2 \cdot 10^{-12} \cdot (T/298.0)^{1.1}$ & \\
    \hline
    T1 & O$_3$ + M $\rightarrow$ O$_2$ + O + M & $7.16 \cdot 10^{-10} \cdot e^{-11200.0/T} \cdot N$ & 300 - 3000 & NIST \\
    T2 & HO$_2$ + M $\rightarrow$ O$_2$ + H + M & $2.41 \cdot 10^{-8} \cdot (T/298.0)^{-1.18} \cdot e^{-24415.0/T} \cdot N$ & 200 - 2000 & NIST \\
    T3 & H$_2$O$_2$ + M $\rightarrow$ OH + OH + M & $2.01 \cdot 10^{-7} \cdot e^{-22852.0/T} \cdot N$ & 700 - 1500 & NIST \\
    \hline
P1	&	H$_2$O + h$\nu$ $\rightarrow$ H + OH	&	0.89 (100 - 144 nm)	&	298	&	\citet{burkholder2015}	\\
	&		&	1 (145 - 198 nm)	&	298	&	\citet{burkholder2015}	\\
P2	&	H$_2$O + h$\nu$ $\rightarrow$ H$_2$ + O($^1$D)	&	0.11 (100 - 144 nm)	&	298	&	\citet{burkholder2015}	\\
P3	&	HO$_2$ + h$\nu$ $\rightarrow$ OH + O	&	1 (185 - 260 nm)	&	298	&	\citet{burkholder2015}	\\
    \enddata
    \tablecomments{The unit of the temperature, $T$, is K and the unit of the number density, $N$, is cm$^{-3}$. References with * are wavelength and temperature dependent parametrizations of the quantum yields.\\
    (This table is available in its entirety in a machine-readable form in the online journal. A portion is shown here for guidance regarding its form and content.)}
\end{deluxetable*}

\begin{deluxetable}{lccc} \label{tab:xs}
    \tabletypesize{\footnotesize}
    \tablecolumns{8}
    \tablecaption{Cross sections used in the photochemical module. The unit of the wavelengths range is nm and the unit of the temperature range is K.}
    \tablehead{Specie & \colhead{Wavelength}  & \colhead{Temperature} & \colhead{Reference}}  
    \startdata
    O$_2$	&	100 - 113	&	298	&	\citet{Brion1979}	\\
    	&	115 - 179	&	298	&	\citet{Lu2010}	\\
    	&	130 - 175	&	90 - 298	&	\citet{Yoshino2005}	\\
    	&	175 - 205	&	130 - 500	&	\citet{Minschwaner1992}*	\\
    	&	205 - 245	&	90 - 298	&	\citet{burkholder2015}	\\
    	&	245 - 294	&	298	&	\citet{Fally2000}	\\
    O$_3$	&	110 - 186	&	298	&	\citet{Mason1996}	\\
    	&	186 - 213	&	218 - 298	&	\citet{burkholder2015}	\\
    	&	213 - 850	&	193 - 293	&	\citet{Serdyuchenko2014}	\\
    H$_2$O	&	100 - 121	&	298	&	\citet{Chan1993}	\\
    	&	121 - 198	&	298	&	\citet{burkholder2015}	\\
    \enddata
    \tablecomments{References with * are wavelength and temperature dependent parametrizations of the cross sections.\\
    (This table is available in its entirety in a machine-readable form in the online journal. A portion is shown here for guidance regarding its form and content.)}
\end{deluxetable}

We use BLACKWOLF (BerLin Atmospheric Chemical Kinetics and photochemistry module With application to exOpLanet Findings) to calculate the atmospheric composition profiles of terrestrial planets. BLACKWOLF is based on previous photochemistry module versions \citep{pavlov2002,rauer2011,gebauer2018} which have been used for multiple studies in our department \citep[e.g.][]{grenfell2013, grenfell2014,scheucher2018,wunderlich2019}. 

The chemical reactions network of BLACKWOLF is fully flexible in the sense that chemical species and reactions can be easily added or removed. Further, the network  can be adapted depending on e.g. the main composition, temperature or surface pressure of the planetary atmosphere in question. The full network consists of 1127 reactions for 128 species, including 832 bi-molecular reactions, 117 termolecular reactions, 53 thermo-dissociation reactions and 125 photolysis reactions. It was developed to compute N$_2$, CO$_2$, H$_2$ and H$_2$O-dominated atmospheres of terrestrial planets orbiting a range of host stars. The network does not include all forward and backward reactions to consistently simulate equilibrium chemistry for high pressure and high temperature regimes. Hence, we limit the usage of the photochemical module to pressures below 100~bar and temperatures below 800~K. Details of the kinetic reactions can be found in Section~\ref{app:kinetics}. 

We consider photochemical reactions for 81 absorbers using wavelength and temperature dependent cross sections. The wavelength and temperature coverage with the corresponding references of all quantum yields and cross sections are given in Table~\ref{tab:reactions} and Table~\ref{tab:xs}. All wavelength dependent data is binned to 133 bands between 100 and 850 nm. See Section~\ref{app:xs} for more details on the selection, binning and interpolation of cross section and quantum yield data. 
For the two-stream radiative transfer, based on \citet{toon1989}, we consider 81 absorbers and the same eight Rayleigh scatterers as in the climate module \citep{shardanand1977,allen1973}. 

The model considers upper and lower boundary conditions of each chemical specie. At the upper boundary we prescribe atmospheric escape by setting either a fixed flux $\Phi_{\text{TOA}}$ in molecules cm$^{-2}$ s$^{-1}$ or an effusion velocity $\nu_{\text{eff}}$ in cm s$^{-1}$. We calculate the molecular diffusion coefficients for the diffusion-limited escape velocity of H and H$_2$ in N$_2$, CO$_2$ or H$_2$-dominated atmospheres from the parametrization shown in \citet{hu2012}. This was derived from the gas kinetic theory and the coefficients are obtained by fitting to experimental data from \citet{marrero1972} and \citet{banks1973}.
Following the upper limit of \citet{luger2015} we assume that the oxygen escape flux is one-half the hydrogen escape flux.

The lower model boundary is given by either a fixed volume mixing ratio, $f$, or a net input or loss at the surface, which depends on the deposition velocity, $\nu_{\text{dep}}$ in cm s$^{-1}$, and the surface emission, $\Phi_{\text{BOA}}$ in molecules cm$^{-2}$ s$^{-1}$. The volcanic flux, $\Phi_{\text{VOLC}}$, is distributed over the lower 10~km of the atmosphere. The boundary conditions used for the simulation of the TRAPPIST-1 planetary atmospheres are given in Section~\ref{ssec:scenarios}. Tropospheric lightning emissions of nitrogen oxides, NO$_x$ (NO, NO$_2$), are also included based on the Earth lightning model of \citet{chameides1977}.

To account for the wet deposition of soluble species we use the parametrization of \citet{giorgi1985}. This parametrization takes as input effective Henry's law constants, $H'$, of all soluble species. We use the values of $H'$ published in \citet{giorgi1985} as well as the classical Henry's law constants, $H$, from \citet{sander2015} and consider available parametrizations of the temperature dependence for the solubility.

In a 1D photochemical model the vertical transport can be approximated by eddy diffusion. In previous model versions the eddy diffusion was fixed to a given profile by \citet{massie1981}, which approximates Earth's vertical mixing. BLACKWOLF uses a parametrization of the eddy diffusion coefficient, similar to \citet{gao2015}, which is based on the equations shown in \citet{gierasch1985}. We introduce the parametrization and compare eddy diffusion profiles for Earth, Venus and Mars in Section~\ref{app:eddy}. \ \\

\subsubsection{Chemical kinetics} \label{app:kinetics}
The chemical network used in previous studies such as \citet{grenfell2007,rauer2011,grenfell2013,wunderlich2019} is based on \citet{kasting1985}, \citet{pavlov2002} and \citet{segura2003} and is able to reproduce the Earth's atmosphere with an N$_2$-O$_2$-dominated composition. This paper introduces an updated and enhanced network also suitable for CO$_2$ and H$_2$-dominated atmospheres. All species included are listed in Table~\ref{tab:species} and all reactions can be found in the Table~\ref{tab:reactions}. Photochemical reactions are discussed in detail in Section~\ref{app:xs}. The chemical network setup is designed to be fully flexible, meaning that subsets of species or reactions can be chosen. 

A large number of chemical reactions are taken from the network presented in \citet{hu2012}. Since we focus on the atmosphere of terrestrial planets in the habitable zone around their host stars, we do not include reactions which are only valid at temperatures above 800~K. From the network of \citet{hu2012} we do not include reactions with hydrocarbon molecules that have more than two carbon atoms. For higher hydrocarbon chemistry we include the reactions up to C$_5$ shown in \citet{arney2016}. This network has been used and validated in multiple studies focusing on the influence of hydrocarbon haze production on atmospheric composition and climate for a range of different atmospheric conditions \citep[e.g.][]{arney2016,arney2017,arney2018}.

Furthermore we update the chlorine chemistry for Earth-like atmospheres with the reaction coefficients from \citet{burkholder2015} and add new reactions, taken from the online database of the National Institute of Standards and Technology \citep[NIST\footnote{http://kinetics.nist.gov},][]{mallard1994}. In particular we include reactions which are important for the destruction and build-up of chloromethane (CH$_3$Cl) for Earth-like atmospheres. Further, we include chlorine and sulphur chemical reactions known to be relevant in CO$_2$-dominated atmospheres such as Mars and Venus from \citet{zhang2012}. Following e.g. \citet{zahnle2008} we multiply all termolecular reaction rates by a bathgas factor of 2.5 when CO$_2$ is the main constituent of the atmosphere and is therefore acting as third body in the termolecular reactions. 

If multiple references are found for the same reaction we compare the reaction rates assuming a temperature of 288~K and decide case by case which reaction rate is considered. If the rates do not differ by more than a factor of three, we use the reference which considers a temperature dependence. If non or multiple rates include a temperature dependence we use the reaction rate from the most recent reference.
For reaction rates which differ significantly from each other we choose the rate which is in agreement with the rates listed in the NIST database. 

To validate that BLACKWOLF is able to simulate the photochemistry of CO$_2$-dominated atmospheres we model the atmospheres of modern Mars and modern Venus above the cloudtop and compare the results with observations (see Section~\ref{app:validation}). 


\subsubsection{Cross sections and quantum yields} \label{app:xs}

The cross section data are taken from the MPI Mainz Spectral Atlas \citep{keller2013}, the JPL Publication No. 15-10 \citep{burkholder2015}, \citet{mills1998} and \citet{zahnle2008}.
In the case that there are multiple cross section data available with the same wavelength and temperature coverage, we follow the recommendations of the JPL Chemical Kinetics and Photochemical Data Publication No. 15-10 \citep{burkholder2015}. If no recommendation was given, we decided case by case which data to use, depending on the consistency of the data with other publications, the year of publication, temperature coverage and wavelength resolution. The quantum yields of the photochemical reactions are taken from \citet{burkholder2015,hu2012,mills1998} and the MPI Mainz Spectral Atlas \citep{keller2013}. The wavelength and temperature range with the corresponding references of all quantum yields and cross sections are given in Table~\ref{tab:reactions} and Table~\ref{tab:xs}. 

For cases with a wavelength gap between two datasets we set the cross sections to zero within the gap. We also assume the cross sections to be zero for wavelengths longer or shorter than covered by the available datasets. Quantum yields are interpolated between different datasets. Further, the quantum yields are extrapolated to 100~nm, the lower wavelength limit of the model, and up to the wavelength which corresponds to the bond energy of the reaction stated in \citet{burkholder2015}. Temperature dependent cross sections and quantum yields are interpolated linearly to the temperature of the atmospheric level. 

\subsubsection{Eddy diffusion} \label{app:eddy}

\picfullwide{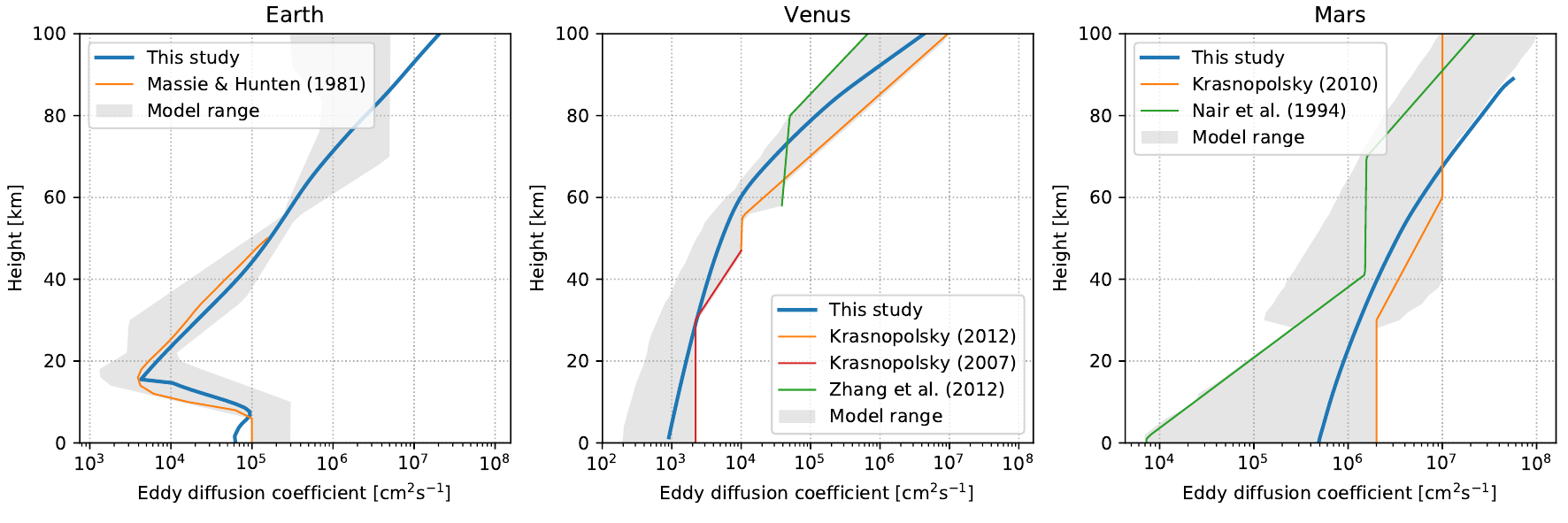}{Profiles of eddy diffusion coefficients, $K$ in cm$^2$s$^{-1}$ for modern Earth (left panel), Venus (middle panel) and Mars (right panel) calculated with Eq.~(\ref{eq:edd}) shown in blue. The $K$ profile of Earth derived from trace gases by \citet{massie1981} is shown in orange. Assumed profiles for Mars in orange from \citet{krasnopolsky2010} and in green from \citet{nair1994}. Assumed profiles for Venus are shown in orange from \citet{krasnopolsky2012}, in red from \citet{krasnopolsky2007} and in green from \citet{zhang2012}. 
Gray shading indicates range of $K$ for multiple model studies (see text for details).}

The eddy diffusion coefficient, $K$, in cm$^2$ s$^{-1}$ as a function of altitude is assumed analogous to that for heat as derived for free convection by \citet{gierasch1985}: 

\begin{equation} \label{eq:edd} 
 K =  \frac{H}{3}\left( \frac{L}{H} \right)^{4/3} \left( \frac{R\sigma T^{4}}{\mu \rho C_{\text{p}}} \right)^{1/3},
\end{equation}
where $H$ is the scale height, $R$ is the universal gas constant, $\sigma$ is the Stefan-Boltzmann constant, $\mu$ is the atmospheric molecular weight, $\rho$ is the atmospheric density, $C_{\text{p}}$ is the atmospheric heat capacity, and $L$ is the mixing length. 

Equation~(\ref{eq:edd}) was also used by e.g. \citet{ackerman2001} and \citet{gao2015} to estimate $K$. 
To fit the $K$ profile of Earth, Mars and Venus we adapt the formula for $L$, which was introduced by \citet{ackerman2001}:

\begin{equation} \label{eq:lmix} 
 L = 
   \begin{cases} 
     H \cdot \text{max}(0.1,\Gamma/\Gamma_{\text{ad}}) & z < z_{\text{ct}} \\ 
     \frac{H_{\text{ct}}}{4} \left(\frac{20}{p_{\text{0}}} + \left(\frac{1}{p}\right)^{1/4}\right)   & z \geq z_{\text{ct}} 
  \end{cases} ,
\end{equation}
where $\Gamma$ is the atmospheric lapse rate, $\Gamma_{ad}$ is the adiabatic lapse rate, $p$ is the atmospheric pressure, $p_{\text{0}}$ is the surface pressure, $z_{\text{ct}}$ is the height of the cold trap and $H_{\text{ct}}$ is the scale height at $z_{\text{ct}}$. 

For a planet with an ocean, such as Earth, $z_{\text{ct}}$ is the atmospheric layer where water condenses out, i.e. at the lowest layer where $\frac{p_{\text{sat}}}{p}$ starts to increase with height. $p_{\text{sat}}$ is the saturation pressure of water. For a planet without an ocean, such as Mars and Venus, the eddy diffusion can be well described by breaking gravity waves alone \citep[see e.g.][]{izakov2001} and $z_{\text{ct}}$ is set to 0~m. 

The left panel of Figure~\ref{fig:Eddy} shows the calculated $K$ profile for Earth compared to the $K$ profile derived from trace gases by \citet{massie1981}. The gray shaded region represents a range of observational fits from multiple models \citep{wofsy1972,hunten1975,allen1981}. The parametrized values match well the results shown in \citet{massie1981} and lie within the model range except close to the surface, where surface properties can influence transport and towards the upper mesosphere, where e.g. gravity wave breaking can influence mixing and energy budgets. We do not consider a constant eddy diffusion coefficient profile for Earth in the mesosphere and thermosphere as proposed by e.g. \citet{allen1981} in order to enable the calculation of $K$ to be as general as possible without further assumptions. For most planets $K$ is found to increase towards high altitudes \citep[see e.g.][]{zhang2018}. Note that the model also has the possibility to use a fixed, predefined $K$ profile.

The middle panel of Figure~\ref{fig:Eddy} shows reasonable agreement for the calculated $K$ profile of Venus with the assumed profiles from \citet{krasnopolsky2007}, \citet{krasnopolsky2012} and \citet{zhang2012}. The maximum values of these three studies represent the upper limit of the model range. The lower limit of the model range is taken from \citet{izakov2001}.  

The calculated $K$ profile for the Martian atmosphere, compared to the assumed profiles from \citet{krasnopolsky2010} and \citet{nair1994}, are shown in the right panel of Figure~\ref{fig:Eddy}. The lower limit of the model range is from \citet{nair1994} up to 30~km and from \citet{montmessin2017} thereabove. The upper limit is from \citet{krasnopolsky2010} and \citet{krasnopolsky2006}. 
Figure~\ref{fig:Eddy} shows that the Eq.~(\ref{eq:edd}) can represent well the $K$ profiles of Earth, Mars and Venus and hence, is suitable to apply to the scenarios we consider for the TRAPPIST-1 planets.
\ \\ \ \\ \ \\
\subsection{Model validation} \label{app:validation}
\subsubsection{Earth} \label{sapp:earth}

\begin{deluxetable*}{lccccccc} \label{tab:Earth_flux}
    \tabletypesize{\footnotesize}
    \tablecolumns{8}
    \tablecaption{$\Phi_{\text{BOA}}$ and $\Phi_{\text{VOLC}}$ of the Earth in molecules~cm$^{-2}$~s$^{-1}$. }
    \tablehead{Specie & \colhead{Anthropogenic}  & \colhead{Ref.} & \colhead{Biogenic} & \colhead{Ref.} & \colhead{Volcanic} & \colhead{Ref.} & \colhead{Biogenic and Volcanic}}  
    \startdata
  O$_2$      &  -                    &  -   & 1.21$\cdot$10$^{12}$ & calc. & -                    & -    &  1.21$\cdot$10$^{12}$ \\
  CH$_4$     &  7.70$\cdot$10$^{10}$ & (1)  & 6.30$\cdot$10$^{10}$ & (1)   & 1.12$\cdot$10$^{8}$  & (2)  &  6.31$\cdot$10$^{10}$ \\
  CO         &  1.16$\cdot$10$^{11}$ & (3)  & 1.07$\cdot$10$^{11}$ & (3)   & 3.74$\cdot$10$^{8}$  & (2)  &  1.07$\cdot$10$^{11}$ \\
  N$_2$O     &  6.58$\cdot$10$^{8}$  & (4)  & 7.80$\cdot$10$^{8}$  & (4)   & -                    & -    &  7.80$\cdot$10$^{8}$  \\
  NO         &  2.46$\cdot$10$^{9}$  & (4)  & 3.38$\cdot$10$^{8}$  & (4)   & -                    & -    &  3.38$\cdot$10$^{8}$  \\
  H$_2$S     &  1.97$\cdot$10$^{7}$  & (5)  & 1.84$\cdot$10$^{9}$  & (5)   & 1.89$\cdot$10$^{9}$  & (2)  &  3.73$\cdot$10$^{9}$  \\
  SO$_2$     &  1.70$\cdot$10$^{10}$ & (5)  & -                    & -     & 1.34$\cdot$10$^{10}$ & (2)  &  1.34$\cdot$10$^{10}$ \\
  NH$_3$     &  3.57$\cdot$10$^{9}$  & (6)  & 8.15$\cdot$10$^{8}$  & (6)   & -                    & -    &  8.15$\cdot$10$^{8}$  \\
  OCS        &  4.54$\cdot$10$^{7}$  & (7)  & 1.39$\cdot$10$^{8}$  & (7)   & 2.67$\cdot$10$^{6}$  & (7)  &  1.42$\cdot$10$^{8}$  \\
  HCN        &  1.32$\cdot$10$^{8}$  & (8)  & 1.27$\cdot$10$^{7}$  & (8)   & -                    & -    &  1.27$\cdot$10$^{7}$  \\
  CH$_3$OH   &  2.91$\cdot$10$^{9}$  & (9)  & 3.35$\cdot$10$^{10}$ & (9)   & -                    & -    &  3.35$\cdot$10$^{10}$ \\
  CS$_2$     &  1.15$\cdot$10$^{8}$  & (7)  & 4.98$\cdot$10$^{8}$  & (7)   & 6.23$\cdot$10$^{6}$  & (7)  &  5.05$\cdot$10$^{8}$  \\
  CH$_3$Cl   &  7.97$\cdot$10$^{7}$  & (4) & 1.39$\cdot$10$^{8}$   & (4)  & -                     & -    &  1.39$\cdot$10$^{8}$  \\  
  C$_2$H$_2$ &  9.48$\cdot$10$^{8}$  & (8)  & -                    & -     & -                    & -    &  -                    \\
  C$_2$H$_6$ &  7.09$\cdot$10$^{8}$  & (4) & 8.50$\cdot$10$^{8}$   & (10)  & 5.10$\cdot$10$^{6}$  & (10) &  8.55$\cdot$10$^{8}$  \\
  C$_3$H$_8$ &  5.52$\cdot$10$^{8}$  & (10) & 9.49$\cdot$10$^{8}$  & (10)  & 2.29$\cdot$10$^{6}$  & (10) &  9.51$\cdot$10$^{8}$  \\
  HCl        &  1.32$\cdot$10$^{9}$  & (11) & 5.13$\cdot$10$^{9}$  & (11)  & 4.42$\cdot$10$^{8}$  & (12) &  5.57$\cdot$10$^{9}$  \\
  H$_2$      &  7.43$\cdot$10$^{10}$ & (3)  & 1.86$\cdot$10$^{10}$ & (3)   & 3.75$\cdot$10$^{9}$  & (2)  &  2.23$\cdot$10$^{10}$ \\
    \enddata
    \tablecomments{The biogenic flux of O$_2$ corresponds to the value necessary to reproduce a volume mixing ratio of O$_2$ of 0.21 on modern Earth, assuming a deposition velocity of 1$\cdot$10$^{-8}$cm/s.
    (1) \citet{lelieveld1998};
    (2) \citet{catling2017};
    (3) \citet{hauglustaine1994};
    (4) \citet{seinfeld2016};
    (5) \citet{berresheim1995};
    (6) \citet{bouwman1997};
    (7) \citet{khalil1984};
    (8) \citet{duflot2015};
    (9) \citet{tie2003};
    (10) \citet{etiope2009};
    (11) \citet{legrand2002};
    (12) \citet{pyle2009}}
\end{deluxetable*}

\begin{deluxetable}{lccc} \label{tab:Earth_vdep}
    \tabletypesize{\footnotesize}
    \tablecolumns{3}
    \tablecaption{$\nu_{\text{dep}}$ as measured for the Earth in cm~s$^{-1}$.}
    \tablehead{Specie & \colhead{$\nu_{\text{dep}}$ (cm~s$^{-1}$)}  & \colhead{Reference}}  
    \startdata
  O$_2$        & 1$\cdot$10$^{-8}$     &     \citet{arney2016}     \\
  O$_3$        & 0.4                   &     \citet{hauglustaine1994}     \\
  H$_2$O$_2$   & 1                     &     \citet{hauglustaine1994}     \\
  CO           & 0.03                  &     \citet{hauglustaine1994}     \\
  CH$_4$       & 1.55$\cdot$10$^{-4}$  &     \citet{watson1992}     \\
  NO           & 0.016                 &     \citet{hauglustaine1994}     \\
  NO$_2$       & 0.1                   &     \citet{hauglustaine1994}     \\
  NO$_3$       & 0.1                   &     \citet{hauglustaine1994}     \\
  N$_2$O$_5$   & 4                     &     \citet{hauglustaine1994}     \\
  HNO$_3$      & 4                     &     \citet{hauglustaine1994}     \\
  HO$_2$NO$_2$ & 0.4                   &     \citet{hauglustaine1994}     \\
  SO$_2$       & 1                     &     \citet{sehmel1980}     \\
  NH$_3$       & 1.7075                &     \citet{phillips2004}     \\
  OCS          & 0.01                  &     \citet{seinfeld2016}     \\
  CH$_3$OOH    & 0.25                  &     \citet{hauglustaine1994}     \\
  HCl          & 0.8                   &     \citet{kritz1980}     \\
  HCN          & 0.044                 &     \citet{duflot2015}     \\
  CH$_3$OH     & 1.26                  &     \citet{tie2003}     \\
    \enddata
    \tablecomments{For all other species we use $\nu_{\text{dep}}$ of 0.02~cm~s$^{-1}$, following \citet{hu2012} and \citet{zahnle2008}.}
\end{deluxetable}

\picfullwide{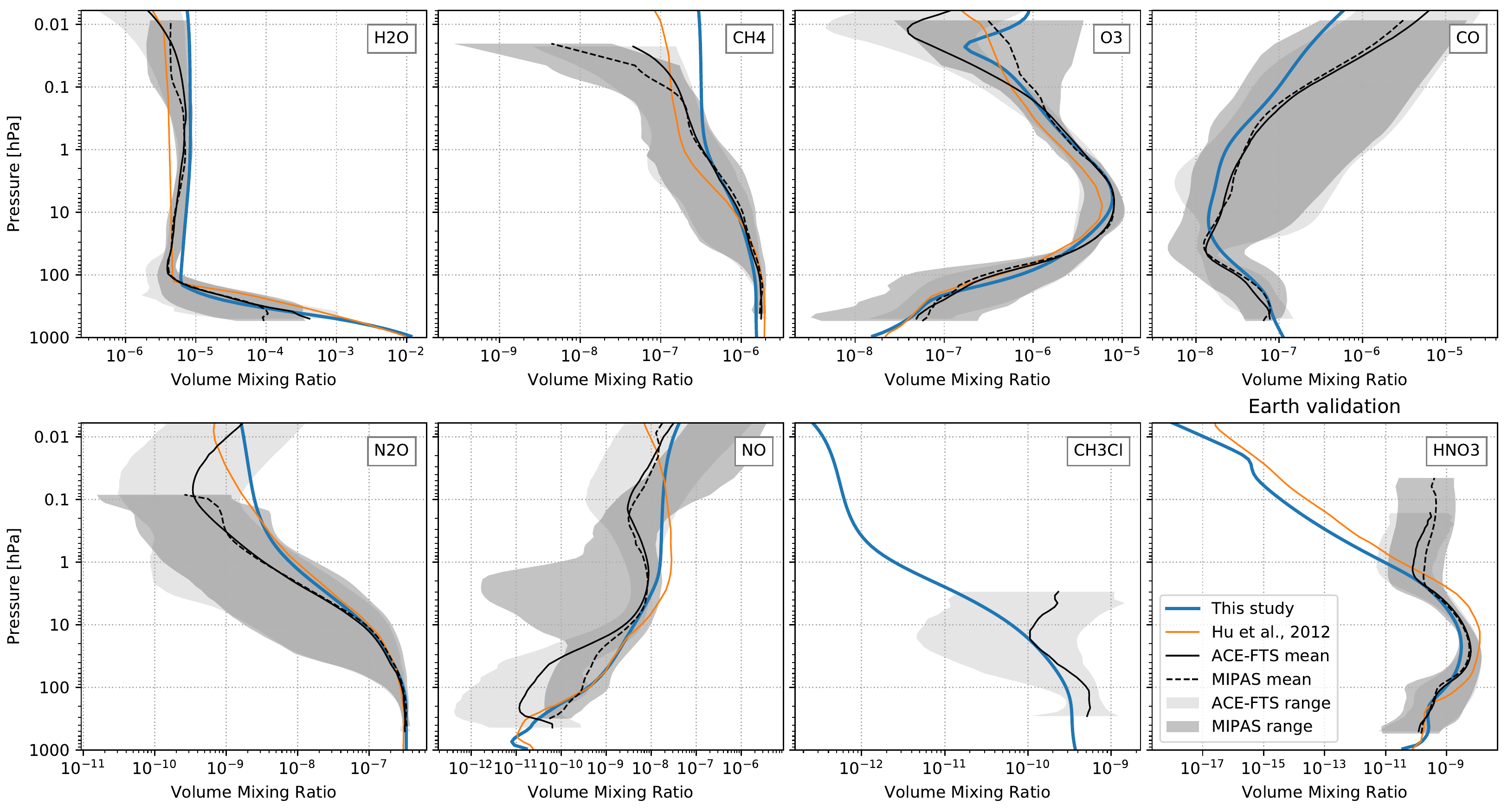}{Earth composition profiles for selected species predicted with our photochemistry model shown in blue, compared to the results from \citet{hu2012} in orange and to MIPAS and ACE-FTS measurements in black. Dark gray shaded regions indicate MIPAS measurement ranges whereas light gray shaded regions indicate ACE-FTS measurement ranges (see text for details).}

We first validate our model by simulating the modern Earth around the Sun and comparing the results with observations from measurements of the Michelson Interferometer for Passive Atmospheric Sounding \citep[MIPAS;][]{fischer08} and the Atmospheric Chemistry Experiment Fourier Transform Spectrometer \citep[ACE-FTS;][]{bernath2017}. Details of the MIPAS and ACE-FTS data processing can be found in \citet{clarmann2009} and \citet{boone2005} respectively. The references of the individual datasets for each species can be found on the MIPAS web page \footnote{www.imk-asf.kit.edu/english/308.php} and the ACE-FTS web page \footnote{ace.scisat.ca/publications/}. 

We select only the data with high quality, determined as following. For MIPAS data we follow the recommendations that the diagonal element of the averaging kernel needs to be at least 0.03 and the visibility flag must be unity \footnote{share.lsdf.kit.edu/imk/asf/sat/mesospheo/data/L3/MIPAS\_L3\_\\ReadMe.pdf}. The ACE-FTS data contains a quality flag indicating physically unrealistic outliers \citep{sheese2015}. The selected data is averaged for each satellite flyover onto a grid with a resolution of 5$^{\circ}$ in latitude by 10$^{\circ}$ in longitude. We repeat this step for each available observation. We take into account 95\% of the data and exclude the 5\% extremes. The maximum and minimum value for each altitude level represents the measured range shown as gray shading in Figure~\ref{fig:Earth}. To calculate the global and annual mean profile of each specie we calculate a monthly mean and from that an annual mean at each grid point. This ensures that each season of the year is equally represented. Finally we average over the grid with a zonal and weighted meridional mean. 

Different to our previous studies we do not tune the surface fluxes to reproduce the observed surface abundances of CO, NO$_2$, CH$_4$ and CH$_3$Cl \citep[e.g.][]{grenfell2013,grenfell2014,wunderlich2019}. Instead we use the sum of observed anthropogenic, biogenic and volcanic surface fluxes (see Table~\ref{tab:Earth_flux}) and observed $\nu_{\text{dep}}$ (see Table~\ref{tab:Earth_vdep}). Also included are modern-day tropospheric lightning emissions of NO$_x$. We apply an upper boundary condition for H and H$_2$ with the parametrization from \citet{hu2012}. 
To simulate modern Earth we use the solar spectrum from \citet{gueymard2004}. The temperature profile simulated with the model is shown in the companion paper \citep{scheucher2020}. To achieve a mean surface temperatures of 288.15~K in our cloud-free model we use a surface albedo of 0.255. 

Figure~\ref{fig:Earth} shows that the photochemistry of the Earth can be reproduced well with the new chemical network. We also compare well to the results shown by \citet{hu2012}. Tropospheric abundances of all shown species lie within the measurement range. In the upper stratosphere and mesosphere the abundances of HNO$_3$ are underestimated in both models compared to measurements. 
This discrepancy could be due to missing NO$_x$-related processes, such as energetic particle precipitation, producing NO$_x$ in the upper mesosphere and subsequent dynamical transport into the stratosphere \citep[see e.g.][]{krivolutsky1993,siskind2000,lopez2005,clilverd2009,funke2005,funke2010,funke2014,funke2016}. 

\subsubsection{Mars} \label{sapp:mars}
As a second validation case we simulate the atmosphere of modern Mars. We use the atmospheric temperature profile from \citet{haberle2017}, representing a scenario with weak dust loading. The data is based on diurnal averages of Mars Climate Sounder (MCS) observations \citep{kleinboehl2009}. The radiative-convective climate module is not used here to calculate the temperature profile since we want to focus on the validation of the photochemistry model. The climate validation for Mars is presented in \citet{scheucher2020}. The mean surface pressure of the reference atmosphere is 5.62~hPa \citep{haberle2017}. We use a bond albedo of 0.25 \citep{williams2010}. The eddy diffusion coefficients are directly calculated in the model (see Section~\ref{app:eddy}). 

In Table~\ref{tab:mars_bnds} we show the boundary conditions used to model the Martian atmosphere. N$_2$ serves as a fill gas and is 2.82\% over the entire atmosphere, which is similar to the measurements of \citet{owen1977} which suggested a volume mixing ratio of 2.7\%.

\begin{deluxetable}{lcccc} \label{tab:mars_bnds}
    \tabletypesize{\footnotesize}
    \tablecolumns{5}
    \tablecaption{Boundary conditions of modern Mars.}
    \tablehead{Specie & Lower & Ref. & Upper  & Ref. }
    \startdata
  CO$_2$   & $f$ = 0.9532             & (1) & $\Phi_{\text{TOA}}$ = 0  &  -    \\
  H$_2$O   & $f$ = 3$\cdot$10$^{-4}$  & (1)   & $\Phi_{\text{TOA}}$ = 0  &  -    \\
  CH$_4$   & $\Phi_{\text{BOA}}$ = 7.5$\cdot$10$^{3}$ & (2)   & $\Phi_{\text{TOA}}$ = 0  &  -    \\
  SO$_2$   & $\Phi_{\text{BOA}}$ = 1.5$\cdot$10$^{6}$  & (3)   & $\Phi_{\text{TOA}}$ = 0  &  -    \\
  HCl      & $\Phi_{\text{BOA}}$ = 2.4$\cdot$10$^{4}$  & (4)   & $\Phi_{\text{TOA}}$ = 0  &  -    \\
  H$_2$    & $\Phi_{\text{BOA}}$ = 0  & - & $\nu_{\text{eff}}$ = 3.39  &  (5)    \\
  H        & $\Phi_{\text{BOA}}$ = 0  & - & $\nu_{\text{eff}}$ = 3080   &  (6)    \\
  O        & $\Phi_{\text{BOA}}$ = 0  & - & $\Phi_{\text{TOA}}$ = 1$\cdot$10$^{7}$    &  (7)    \\
  O$_2$       & $\nu_{\text{dep}}$ = 1$\cdot$10$^{-8}$  & (8) & $\Phi_{\text{TOA}}$ = 0     &  -    \\
  CO       & $\nu_{\text{dep}}$ = 1$\cdot$10$^{-8}$  & (9) & $\Phi_{\text{TOA}}$ = 0     &  -   \\
  other    & $\nu_{\text{dep}}$ = 2$\cdot$10$^{-2}$  & (7) & $\Phi_{\text{TOA}}$ = 0     &  -   \\
    \enddata
    \tablecomments{See Section~\ref{ssec:photo_model} for description of how the boundaries are included in the model. $\Phi_{\text{BOA}}$ and $\Phi_{\text{TOA}}$ are in molecules cm$^{-2}$ s$^{-1}$, $\nu_{\text{dep}}$ and $\nu_{\text{eff}}$ are in cm~s$^{-1}$. Following \citet{zahnle2008}, for all species not listed here we assume a $\nu_{\text{dep}}$ of 0.02~cm~s$^{-1}$. (1)~\citet{owen1977}, (2)~$\Phi_{\text{BOA}}$ necessary to fit the mean surface value of $f_{\text{CH$_4$}}$~=~4$\cdot$10$^{-10}$ \citep{webster2018}, (3) $\Phi_{\text{BOA}}$ necessary to fit the upper limit of $f_{\text{SO$_2$}}$~=~3$\cdot$10$^{-10}$ \citep{encrenaz2011}, (4)~$\Phi_{\text{BOA}}$ necessary to fit the upper limit of $f_{\text{HCl}}$~=~2$\cdot$10$^{-10}$ \citep{hartogh2010}, (5)~$\nu_{\text{eff}}$ necessary to fit $f_{\text{H$_2$}}$~=~1.5$\cdot$10$^{-5}$ at TOA \citep{krasnopolsky2001}; \citet{nair1994} used $\nu_{\text{eff}}$ = 33.9~ cm~s$^{-1}$, (6)~\citet{nair1994}, (7)~\citet{zahnle2008}, (8)~\citet{arney2016}, (9)~\citet{kharecha2005}.  We use a constant volume mixing ratio of argon profile of 1.6\% \citep{owen1977}. N$_2$ serves as a fillgas.
}
\end{deluxetable}

\picfullwide{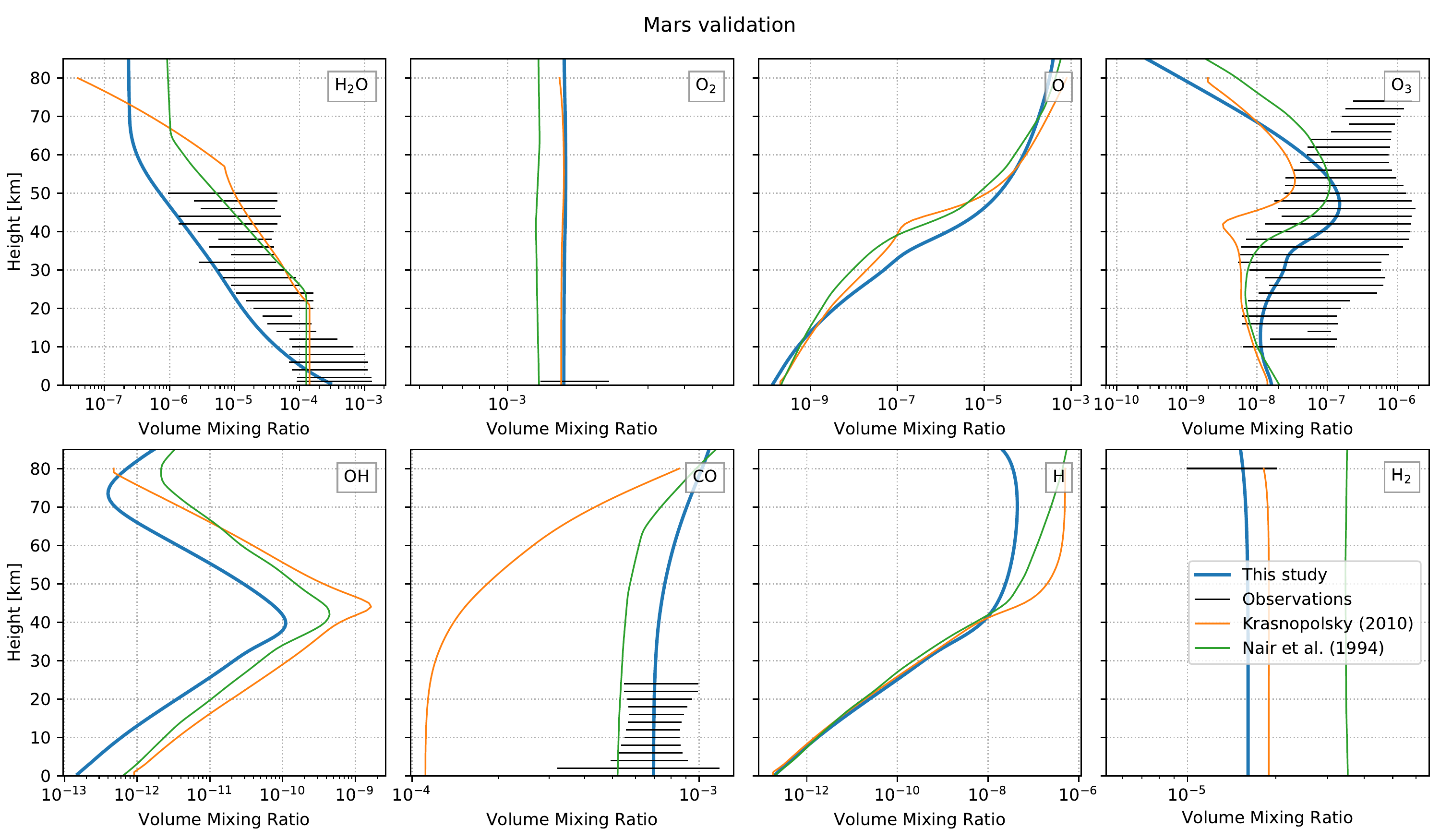}{Mars composition profiles for selected species predicted with our photochemistry model shown in blue, compared to the results from \citet{krasnopolsky2010} in orange, \citet{nair1994} in green and a range of multiple observations in black (see text for details).}

Figure~\ref{fig:Mars} shows the profile of selected atmospheric species compared to the model results of \citet{krasnopolsky2010} and the following measurements. For H$_2$O we take into account Mars Express PFS (Planetary Fourier Spectrometer) nadir measurements up to 30~km from \citet{montmessin2019} and SPICAM (Spectroscopy for the Investigation of the Characteristics of the Atmosphere of Mars) measurements above 20~km from \citet{fedorova2009}. O$_3$ ranges are taken from nighttime and sunrise/sunset measurements \citep{montmessin2013,lebonnois2006}. CO observational ranges are taken from retrieval uncertainties around 800~ppm from PFS/Mars Express infrared nadir observations \citep{bouche2019}. The H$_2$ range at 80~km is given in \citet{krasnopolsky2001} and O$_2$ range at the surface is taken from \citet{trainer2019}. We compute the observational ranges by finding the lowest and highest value in a 2~km grid from measured profiles or observations of the mixing ratio at a given altitude. Note that surface values are located at 1~km for visibility purposes. 

The Martian atmosphere simulated with the photochemistry model compares well with the results from \citet{krasnopolsky2010} and \citet{nair1994}. The model simulates H$_2$O abundances close to the lower minimum of measured concentrations. 
When using an eddy diffusion flux increased by a factor of ten, more water is transported upwards and the modelled H$_2$O abundances fit to the measurements (not shown). Since we model an aerosol free atmosphere the low H$_2$O content is consistent with observations of \citet{vandaele2019} showing increased atmospheric H$_2$O during dust storms. Note that \citet{krasnopolsky2010} and \citet{nair1994} used a predefined H$_2$O profile while we calculate the H$_2$O profile consistently in the photochemical model. 
The underestimation of the O$_3$ content above 60~km may be related to diurnal changes in the solar zenith angle, not included in the model. We obtain a surface O$_2$ concentration of 1552~ppm which is consistent with the global mean of 1560$\pm$54~ppm inferred by \citet{krasnopolsky2017} and also in the range of the seasonal variation of O$_2$ \citep[1300~-~2200~ppm,][]{trainer2019}. 

In summary we show that our photochemistry model gives consistent results compared to previous photochemistry models and observations of the Martian atmosphere. Different from many previous models, we also simulate consistently the chemistry of chlorine, sulphur and methane. The emission fluxes required to reproduce observations of CH$_4$, HCl and SO$_2$ are shown in Table~\ref{tab:mars_bnds}. The Martian CH$_4$ chemistry will be discussed in detail in a follow up paper by Grenfell et al. (in prep).

\subsubsection{Venus} \label{sapp:venus}
\begin{deluxetable}{lcc} \label{tab:venus_bnds}
    \tabletypesize{\footnotesize}
    \tablecolumns{3}
    \tablecaption{Boundary conditions of modern Venus.}
    \tablehead{Specie & Lower & Ref.}
    \startdata
  CO$_2$   & $f$ = 0.965             & \citet{zhang2012}   \\
  CO       & $\nu_{\text{m}} = 0.1 K/H$   & \citet{krasnopolsky2012}  \\
  H2O      & $f$ = 4.0$\cdot$10$^{-6}$  & tuned     \\
  OCS      & $f$ = 1.2$\cdot$10$^{-8}$  & tuned      \\
  NO       & $f$ = 5.5$\cdot$10$^{-9}$  & \citet{zhang2012}      \\
  HCl      & $f$ = 1$\cdot$10$^{-6}$ & tuned (calc. edd. diff.)    \\
  HCl      & $f$ = 4$\cdot$10$^{-7}$ &\citet{zhang2012} (K12 edd. diff.)   \\
  SO2      & $f$ = 3.5$\cdot$10$^{-6}$  & \citet{zhang2012}       \\
  other    & $\nu_{\text{m}} = K/H$   & \citet{zhang2012}      \\
    \enddata
    \tablecomments{For all species not listed here we assume a maximum deposition velocity $\nu_{\text{m}} = K/H$, using $K$ and $H$ at 58~km to take into account that our BoA is not the surface \citep[see][]{zhang2012,krasnopolsky2012}. $f_{\text{HCl}}$ = 1$\cdot$10$^{-6}$ for the run with a calculated $K$ and $f_{\text{HCl}}$ = 4$\cdot$10$^{-7}$ for the run with $K$ taken from \citet{krasnopolsky2012}. N$_2$ serves as fillgas.
}
\end{deluxetable}

\picfullwide{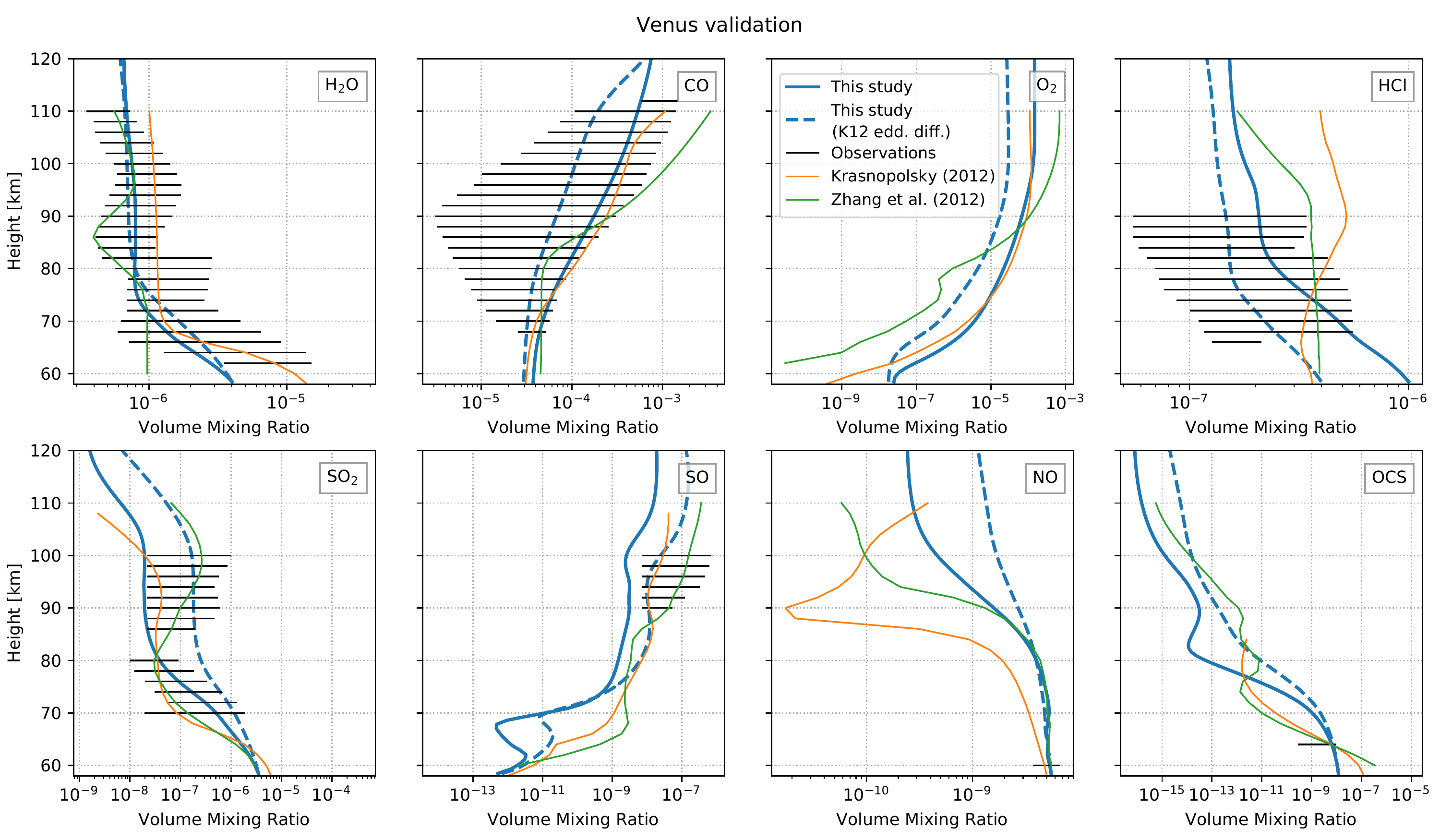}{Venus composition profiles for selected species predicted with our photochemistry model with calculated $K$ (solid blue line) and with $K$ taken from \citet{krasnopolsky2012} with breakpoint $h_\text{e}$ at 65~km (K12 edd. diff., dashed blue line), compared to the results from \citet{krasnopolsky2012}, \citet{zhang2012} and a range of observations inferred from multiple studies (see text for details).}

Predicting the atmospheric composition of Venus is challenging since details of the sulphur chemistry are not understood completely \citep[e.g.][]{mills2007,zhang2012,vandaele2017}. The atmospheric chemistry of Venus below and above the cloud deck is usually modeled separately. We validate our model by calculating the atmosphere of Venus only in the photochemical regime above the cloud top at $\sim$58~km, where direct observations of chemical species are available. The temperature profile is taken from the Venus International Reference Atmosphere VIRA-1 \citep{seiff1985}.

The boundary conditions are presented in Table~\ref{tab:venus_bnds}. Following \citet{zhang2012} and \citet{krasnopolsky2012} we use fixed volume mixing ratios at BoA for key species to fit the observed values and we assume a downward flux of all other species depending on $K$ and $H$ (see also Section~\ref{app:eddy}). Figure \ref{fig:Venus} shows the profiles of the species with existing observations and profiles taken from \citet{zhang2012} and \citet{krasnopolsky2012}. 

The range of observational values  is derived by combining multiple studies. The H$_2$O range is generated by combining measurements from \citet{bertaux2007} and measurements shown in Figure~3 of \citet{krasnopolsky2012}. CO measurements are taken from \citet{svedhem2007} and Figure~2 of \citet{krasnopolsky2012}. HCl measurements are taken from \citet{sandor2012} and \citet{bertaux2007}. For the observational range of SO$_2$ and SO we use Venus Express solar occultations in the infrared range and SPICAV (Spectroscopy for Investigation of Characteristics of the Atmosphere of Venus) occultations from \citet{belyaev2012} and submillimeter measurements from \citet{sandor2010}. The OCS observation is taken from \citet{krasnopolky2010b} and NO measurements from \citet{krasnopolsky2006}. As for the Mars validation we compute the observational ranges by finding the lowest and highest value in a 2~km grid. 

We find that our model is able to reproduce the Venus atmosphere above 58~km and leads to broadly comparable results as for other photochemical models. Our model reproduces the measurements best with a H$_2$O mixing ratio of 4.0$\cdot$10$^{-6}$, which is in between the values shown in \citet{krasnopolsky2012} and \citet{zhang2012}. 
The HCl profile of our model is consistent with the decrease between 70 and 100~km found by observations \citep{sandor2012} and was not reproduced by the models of \citet{krasnopolsky2012} and \citet{zhang2012}.  
On using our calculated eddy diffusion coefficients we underestimate the abundances of SO$_2$ and SO between 90 and 100~km. Using larger eddy diffusion coefficients from \citet{krasnopolsky2012} we then lie in the observational range of SO$_2$ and SO between 90 and 100~km but slightly overestimate the SO$_2$ abundances around 80~km. This degeneracy may be caused by the missing consideration of sulphur hazes in the upper atmosphere \citep[see e.g.][]{gao2014}.

In summary we find that we can predict the upper atmosphere of Venus similarly well as other models, even without consideration of the effect of hazes above the cloud layer. 

\subsection{Transmission spectra} \label{ssec:garlic}
The climate-photochemistry model is used to simulate atmospheric temperature and composition profiles of potential atmospheres of TRAPPIST-1~e and \hbox{TRAPPIST-1~f}. With the resulting profiles we produce transmission spectra of the planetary atmospheres using the "Generic Atmospheric Radiation Line-by-line Infrared Code" \citep[GARLIC;][]{schreier2014, schreier2018}.
GARLIC has been used in recent exoplanet studies such as \citet{scheucher2018,katyal2019,wunderlich2019}.

We simulate transmission spectra including 28 atmospheric species\footnote{OH, HO$_2$, H$_2$O$_2$, H$_2$CO, H$_2$O, H$_2$, O$_3$, CH$_4$, CO, N$_2$O, NO, NO$_2$, HNO$_3$, ClO, CH$_3$Cl, HOCl, HCl, ClONO$_2$, H$_2$S, SO$_2$, O$_2$, CO$_2$, N$_2$, C$_2$H$_2$, C$_2$H$_4$, C$_2$H$_6$, NH$_3$, HCN} between 0.4~$\mu$m and 12~$\mu$m. Line parameters are taken from the HITRAN 2016 database \citep{gordon2017} and the Clough-Kneizys-Davies (CKD) continuum model \citep{clough1989}. Additionally Rayleigh extinction is considered \citep{murphy1977,clough1989, sneep2005, marcq2011}. In the visible we use the cross sections at room temperature (298~K) for O$_3$, NO$_2$, NO$_3$ and HOCl listed in Table~\ref{tab:xs}. 

For the 1D climate-photochemistry simulations we do not consider cloud formation. Hence, all the transmission spectra we calculate in this study show cloud-free conditions. However, an Earth-like extinction from uniformly distributed aerosols in the atmosphere can be considered in GARLIC. The aerosol optical depth, $\tau_{\text{A}}$, at wavelength $\lambda$ ($\mu$m) is expressed following \citet{angstrom1929,angstrom1930}: 

\begin{equation} \label{eq:sigma_aer} 
 \tau_{\text{A}} = \beta  \cdot \lambda^{-\alpha},
\end{equation}

\noindent
assuming that the aerosol size distribution follows the Junge distribution \citep{junge1952,junge1955}. 
For the exponent, $\alpha$, we use 1.3, representing the average measured value on Earth \citep[see e.g.][]{angstrom1930,angstrom1961}. The \r{A}ngstr\"{o}m turbidity coefficient, $\beta$, is expressed using the cross section data for the Earth's atmosphere taken from \citet{allen1976}:

\begin{equation} \label{eq:beta}
\beta = 1.4 \cdot 10^{-27}\cdot N_\text{c},
\end{equation}

\noindent
where $N_\text{c}$ is the column density in molecules~cm$^{-2}$ \citep[see also][]{toon1976,kaltenegger2009,yan2015}. 
According to \citet{allen1976} the Eq.~(\ref{eq:beta}) corresponds to clear atmospheric conditions with weak scattering by haze or dust.  

The transmission spectra from GARLIC are expressed as effective heights:

\begin{equation} \label{eq:effhei} 
 h_{\text{e}}(\lambda) =  \int_{0}^{ToA} \Big(1-\mathcal{T}(\lambda,z)\Big)~dz,
\end{equation}

\noindent
 where $\mathcal{T}$ is the transmission along the limb with the tangent altitude, $z$. $h_{\text{e}}$ is the integration over all $\mathcal{T}$ from the surface to the top of atmosphere (ToA) at each wavelength, $\lambda$. The measured transit depth, $t_{\text{depth}}$, of a planet with an atmosphere is the sum of the planet radius, $R_{\text{p}}$, and $h_{\text{e}}$ with respect to the stellar radius, $R_{\text{s}}$.
 The atmospheric transit depth, $t_{\text{atm}}$, only contains the contribution of the atmosphere to the total transit depth:
 
 \begin{equation} \label{eq:tr_depth} 
 t_{\text{atm}}(\lambda) = \frac{(R_{\text{p}} + h_{\text{e}}(\lambda))^2}{R_{\text{s}}^2} - \frac{R_{\text{p}}^2}{R_{\text{s}}^2}.
 \end{equation}

\noindent
In order to detect a spectral feature we make use of the wavelength dependence of $t_{\text{atm}}$. To extract the measurable atmospheric signal, $S_{\text{atm}}$, we subtract the minimum atmospheric transit depth, $t_{\text{min}}$, in the considered wavelength range (baseline) from the $t_{\text{atm}}$ at each wavelength point:

\begin{equation}
  t_{\text{min}} = \min{(t_{\text{atm}}(\lambda))},
\end{equation}
\begin{equation} \label{eq:satm}   
 S_{\text{atm}}(\lambda) = t_{\text{atm}}(\lambda) - t_{\text{min}}.
\end{equation}

\noindent
The wavelength dependent $S_{\text{atm}}$, expressed as parts per million (ppm),  is used to calculate the signal-to-noise ratio (S/N) of molecular features. Taking into account the $h_{\text{e}}(\lambda)$ instead would overestimate the S/N of the spectral features, because that measure would include the continuum extinction.  


\subsection{Signal-to-noise ratio (S/N)} \label{ssec:snr}

\picwide{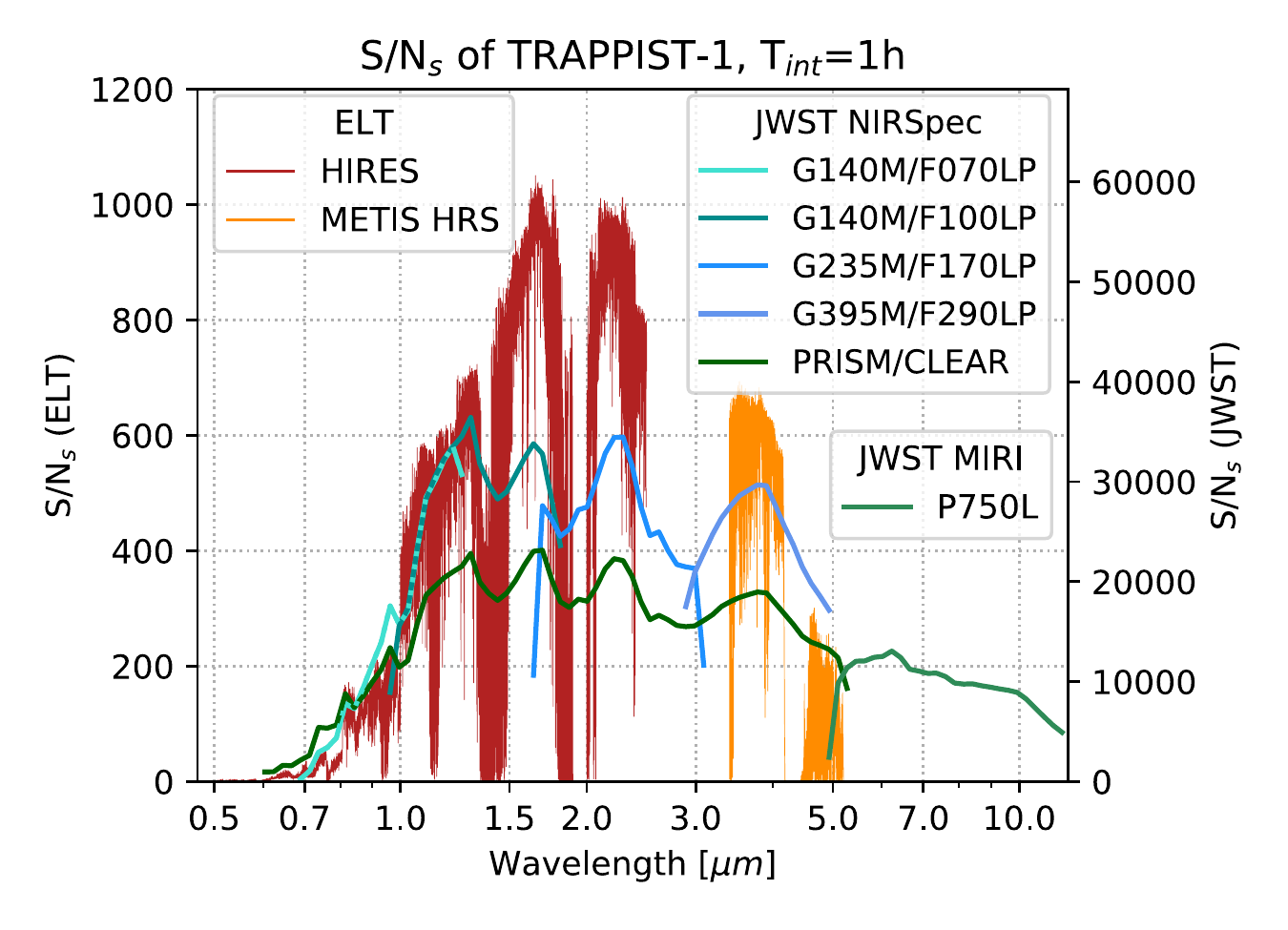}{Stellar S/N of TRAPPIST-1 for 1~h integration time and binned to a resolving power of $R$=100,000 for ELT (left y-axis) and a $R$=30 for JWST (right y-axis). The conversion factor from the right to the left y-axis is $\sqrt{\frac{100,000}{30}}$, corresponding to a white noise binning of the S/N$_s$. The stellar S/N of JWST is the combination of all NIRSpec filter and disperser and MIRI LRS, calculated with the method presented in \citet{wunderlich2019}. We do not consider a partial saturation strategy as suggested by \citet{batalha2018}. The stellar S/N of ELT is calculated with the ESO ETC Version 6.4.0 \citep{liske2008}.}

\begin{deluxetable*}{llccc} \label{tab:instruments}
    \tabletypesize{\footnotesize}
    \tablecolumns{5}
    \tablecaption{Wavelength coverage and resolving power, $R$, of the instruments on JWST and ELT used to calculate SNR$_s$ of TRAPPIST-1.}
    \tablehead{\colhead{Telescope} & \colhead{Instrument}  & \colhead{Wavelength} & \colhead{$R$} & \colhead{Reference} }  
    \startdata
  JWST        & NIRSpec PRISM/CLEAR    & 0.6~-~5.3~$\mu$m  & $\sim$100  & \citet{birkmann2016b}     \\
  JWST        & NIRSpec G140M/F070LP    & 0.7~-~1.27~$\mu$m  & $\sim$1,000  & \citet{birkmann2016b}     \\
  JWST        & NIRSpec G140M/F100LP    & 0.97~-~1.84~$\mu$m  & $\sim$1,000  & \citet{birkmann2016b}     \\
  JWST        & NIRSpec G235M/F170LP    & 1.66~-~3.07~$\mu$m  & $\sim$1,000  & \citet{birkmann2016b}     \\
  JWST        & NIRSpec G395M/F290LP    & 2.87~-~5.10~$\mu$m  & $\sim$1,000  & \citet{birkmann2016b}     \\
  JWST        & MIRI P750L (LRS)     & 5.0~-~12~$\mu$m  & $\sim$100  & \citet{kendrew2015}     \\
  ELT       & HIRES    & 0.37~-~2.5~$\mu$m  & 100,000  & \citet{marconi2016}     \\
  ELT       & METIS (HRS)   & 2.9~-~5.3~$\mu$m  & 100,000  & \citet{brandl2016}     \\
    \enddata
\end{deluxetable*}

We determine which atmospheric spectral features of the simulated atmospheres of TRAPPIST-1~e and TRAPPIST-1~f could be detectable with ELT and JWST. 
\citet{lustig-yaeger2019} showed that the S/N for emission spectroscopy of TRAPPIST-1~e and TRAPPIST-1~f is too low to detect spectral features \citep[see also][]{batalha2018}. Hence, we limit our analysis to transmission spectroscopy.

To calculate the S/N of planetary atmospheric feature, S/N$_{\text{atm}}$, of a single transit, we first calculate the S/N of the star, S/N$_{\text{s}}$, integrated over one transit and then multiply this value with $S_{\text{atm}}$:

\begin{equation} \label{eq:snr_atm} 
\text{S/N}_{\text{atm}} = \frac{S_{\text{atm}}}{\sqrt{2}} \cdot \text{S/N}_{\text{s}}.
 \end{equation}
 
The factor $\frac{1}{\sqrt{2}}$ accounts for the fact that the star is observed during in transit and out of transit.
We calculate the number of transits, $n_{\text{tr}}$, necessary to reach an S/N of 5, assuming that all transits improve S/N$_{\text{s}}$ perfectly.
The S/N$_{\text{s}}$ for JWST NIRSpec and MIRI is determined by the method and instrument specifications presented in \citet{wunderlich2019} (see Table~\ref{tab:instruments} for the wavelength coverage and resolving power, $R = \frac{\lambda}{\Delta \lambda}$). 

The S/N$_{\text{s}}$ of the ELT High Resolution Spectrograph \citep[HIRES;][]{marconi2016} is calculated with the ESO Exposure Time Calculator\footnote{https://www.eso.org/observing/etc/bin/gen/form?INS.NAME=\\E-ELT+INS.MODE=swspectr} (ETC) Version 6.4.0 from November 2019 \citep[see updated documentation\footnote{https://www.eso.org/observing/etc/doc/elt/etc\_spec\_model.pdf} from][]{liske2008}. 
The ETC uses the background sky model\footnote{https://www.eso.org/sci/facilities/eelt/science/drm/tech\_data/\\background/} for the Cerro Paranal and considers photon and as well as detector noises such as readout noise and dark current. The ETC assumes a spectrograph with a throughput of 25\%, independent of the resolving power. For HIRES or METIS HRS this value might overestimate the real value. For METIS HRS the expected throughput ranges between 6\% and 21\% (C\'{a}rdenas V\'{a}zquez, personal communication). Hence, we scale down the S/N$_{\text{s}}$ for both instruments to an average throughput of 10\%. 

We assume a telescope with a diameter of 39~m at Paranal in Chile (2,635~m). The planned location of the ELT at Cerro Armazones (3,046~m) is not available in the ETC. The sky conditions are set to a constant airmass of 1.5 and a precipitable water vapour (PWV) of 2.5 \citep{liske2008}. The ETC does not provide the possibility to choose the individual ELT instrumentations but we consider the wavelength coverage and $R$ for the instruments planned for the ELT (see Table \ref{tab:instruments}). 
For each wavelength band we change the radius of the diffraction limited core of the point spread function according to the recommendation in the ETC manual.
The wavelengths from 2.9~$\mu$m to 3.4~$\mu$m cannot be calculated by the current version of the ETC.

To simulate an observation of TRAPPIST-1 we scale the stellar spectrum from \citet{wilson2020} to the J-band magnitude of 11.35 \citep{gillon2016} in order to obtain the input flux distribution. 

The S/N$_{\text{s}}$ for a one hour integration of TRAPPIST-1 for JWST and ELT is shown in Figure~\ref{fig:SNR_T1_1h_JWST_ELT}.
The ground-based facility ELT will have a much larger telescope area compared to the space-borne JWST but its capability of detecting spectral features with low resolution spectroscopy is limited to atmospheric windows with minor telluric contamination. However, high-resolution spectra ($R$ $>$ 25,000) resolve individual lines improving their detectability. The Doppler-shift of the lines during the transit with respect to the absorption lines of the Earth's atmosphere is measurable for close-in planets \citep[see e.g.][]{birkby2018}. Previous theoretical and observational studies have shown that a detection of molecules such as O$_2$, H$_2$O or CO is feasible via cross-correlation \citep[e.g.][]{snellen2013,birkby2013,brogi2018,molliere2019,morales2019,sanchez2019}. 

\picfull{T1_spec_new}{Input stellar spectral energy distribution (SED) of TRAPPIST-1 and the Sun. Red line: TRAPPIST-1 SED with the UV estimated with a semi-empirical model using HST observational data provided by the Mega-MUSCLES survey \citep{wilson2020}, marked W20 SED. Cyan line: TRAPPIST-1 SED with estimated UV flux by scaling the spectrum of Proxima Centauri \citep{lincowski2018}, marked L18 SED. Violet line: TRAPPIST-1 SED with calculated UV flux using a semi-empirical non-LTE model \citet{peacock2019}, marked P19 SED. Black line: solar SED taken from \citet{gueymard2004}. For the FUV/NUV ratio the FUV is integrated between 117-175~nm and the NUV is integrated over 175-320~nm \citep[see][]{tian2014}.}

We adopt a simple approach in order to estimate the number of transits which are necessary to detect e.g. O$_2$, H$_2$O and CO with the cross-correlation method in our simulated atmospheres. We adapted a formula presented in \citet{snellen2015} to calculate the signal-to-noise ratio of the planet, considering the wavelength dependency of $S_{\text{atm}}$ and S/N$_{\text{s}}$

 \begin{equation} \label{eq:cross-corr} 
\text{S/N}_{\text{atm}} = \frac{\sum_{l=0}^{n_{\text{l}}} S_{\text{atm}}(\lambda_{\text{l}}) \cdot S/N_{\text{s}}(\lambda_{\text{l}})}{n_{\text{l}}} \cdot \sqrt{t_{\text{int}}} \cdot \sqrt{n_{\text{l}}},
 \end{equation}

\noindent
where $n_{\text{l}}$ is the number of spectral lines and $t_{\text{int}}$ is the integration time. $t_{\text{int}}$ is calculated by $t_{\text{dur}} \cdot n_{\text{tr}}$, with the transit duration, $t_{\text{dur}}$, and the number of transits, $n_{\text{tr}}$. The S/N$_{\text{s}}$ at the wavelength of the line, $\lambda_{\text{l}}$, used in Eq.~(\ref{eq:cross-corr}), is the S/N$_{\text{s}}$ shifted by one bandwidth to account for the displacement of the spectral line during transit. 

Using Eq.~(\ref{eq:cross-corr}) we find that a 3$\sigma$ detection of O$_2$ on an Earth-twin around an M7 star at a distance of 5~pc might be feasible when co-adding 58 transit observations in the J-band with ELT HIRES, assuming a throughput of 20\%. \citet{rodler2014} suggested that 26 transits are needed to detect O$_2$ when using the same assumptions.

Section~\ref{ssec:speculoos} discusses the detectability of the CO spectral feature in the atmosphere of a hypothetical planets around other low mass stars in the solar neighbourhood. For stars on the Northern sky we calculate the S/N$_{\text{s}}$ for the Thirty Meter Telescope \citep[TMT, ][]{nelson2008}. 
This will have a smaller telescope area than the ELT but will be located at a higher altitude of 4,064~m, compared to 2,635~m at Paranal. Hence, due to the lower PWV and weaker high-altitude turbulence at Mauna Kea the TMT is expected to have a similar performance as the ELT. We compare the S/N$_s$ of ELT with $R$=4,000 at a Vega magnitude of 16 in the J-band to calculation of the S/N$_s$ with the same specifications using the Infrared Imaging Spectrograph (IRIS) on TMT by \citet{wright2014} and find that ELT has a 10\% lower S/N$_s$ than TMT. 

Since the performance of the telecopes during operation is not yet established we simply assume that the TMT provides the same S/N$_s$ as the ELT.


\section{Stellar input and model scenarios} \label{ssec:sed_scenarios}

\subsection{TRAPPIST-1 spectra} \label{ssec:trappist1_spec}
The Spectral Energy Distribution (SED) in the UV has a large impact on the photochemisty of atmospheres of terrestrial planets \citep[see e.g.][]{selsis2002,grenfell2013,grenfell2014,tian2014}. 
In this study we use the semi-empirical model spectrum of TRAPPIST-1 from \citet{wilson2020}, which we will refer to as W20 SED. The constructed SED uses observational data from XMM-Newton for the X-ray regime and from the Hubble Space Telescope (HST) for the 113 to 570~nm range with a gap between 208-279~nm obtained through the Mega-MUSCLES Treasury survey \citep{froning2018}. The wavelengths larger than 570~nm are filled by \citet{wilson2020} with a PHOENIX photospheric model \citep{allard2016,baraffe2015}. 

Figure~\ref{fig:T1_spec_new} compares the Mega-MUSCLES TRAPPIST-1 SED with spectra, presented in previous studies. \citet{lincowski2018} estimated the UV radiation of TRAPPIST-1 by scaling the Proxima Centauri's spectrum to the Ly$\alpha$ measurements of TRAPPIST-1 from \citet{bourrier2017}, in the following referred to as L18 SED. \citet{peacock2019} present a semi-empirical non-local thermodynamic equilibrium (non-LTE) model spectrum of TRAPPIST-1, based on the stellar atmosphere code PHOENIX \citep{hauschildt1993,hauschildt2006,baron2007}, here referred to as P19 SED. 

We bin all spectra into 128 bands for the climate model and 133 bands for the photochemistry model. The spectra for TRAPPIST-1, as well as the solar spectrum from \citet{gueymard2004} are shown in Figure \ref{fig:T1_spec_new}. All SEDs are scaled to an integrated total energy of 1361~W/m$^2$ which is equal to the energy the Earth receives from the Sun. \\

\begin{deluxetable}{lccc} \label{tab:planet_par}
    \tabletypesize{\footnotesize}
    \tablecolumns{4}
    \tablecaption{Planetary parameters used as input for the climate-photochemistry model and to calculate the S/N of spectral features. The planetary radii from \citet{delrez2018b} are corrected according to \citet{kane2018}. The gravity is calculated using given planetary mass and radius.}
    \tablehead{\colhead{Planets} & \colhead{e}  & \colhead{f} & \colhead{Reference}}  
    \startdata
        Radius ($R_{\earth}$)         &   0.94   &   1.08  & \citet{kane2018} \\
        Mass ($M_{\earth}$)           &   0.772   &   0.934 & \citet{grimm2018}  \\
        Gravity (m/s$^2$)             &   8.56    &   7.85 & -  \\
        Irradiation (S$_{\sun}$)      &   0.604  &   0.349 & \citet{delrez2018b}  \\
        Transit duration (min)                 &   55.92  &   63.14 & \citet{delrez2018b}  \\
        Impact parameter $b$ ($R_*$)  &   0.24    &   0.337 & \citet{delrez2018b}    \\
    \enddata
    \tablecomments{Using the updated stellar parameters from \citet{kane2018} the planetary radii are $\sim$3 larger and the gravities $\sim$7 lower than the values used by previous studies such as \citet{lincowski2018}.}
\end{deluxetable}

\begin{deluxetable}{m{0.7cm} m{0.7cm} m{0.7cm} m{0.7cm} m{0.7cm} m{0.7cm} m{0.7cm} m{0.7cm}} \label{tab:tsurf_3d}
    \tabletypesize{\footnotesize}
    \tablecolumns{8}
    \tablecaption{Mean surface temperature predicted with our 1D climate model \citep[see][]{scheucher2020} for different main atmospheric compositions and stellar irradiations of TRAPPIST-1~e and TRAPPIST-1~f (T$_{\text{1D}}$). S$_{\text{D18}}$ corresponds to the irradiation values shown in \citet{delrez2018b} and S$_{\text{G17}}$ corresponds to the values taken from \citet{gillon2017}. The surface temperatures predicted with various 3D models are shown for comparison (T$_{\text{3D}}$). The last column shows the reference of the corresponding 3D model study.}
    \tablehead{\colhead{Planet} & CO$_2$ (bar) & N$_2$ (bar) & CH$_4$ (bar) & T$_{\text{1D}}$ (S$_{\text{D18}}$) & T$_{\text{1D}}$ (S$_{\text{G17}}$) & T$_{\text{3D}}$ (S$_{\text{G17}}$) & \colhead{Ref.}}
    \startdata
    e & 0.01 & 1 & 0 & 253 & 262 & 254 & (1)\\
    e & 0.1 & 1 & 0 & 269 & 279 & 273 & (1) \\
    e & 1 & 1 & 0 & 328 & 337 & 331 & (1) \\
    e & 0 & 1 & 0.01 & 223 & 231 & 211 &(2)  \\
    e & 1 & 0 & 0 & 303 & 312 & 303 &(3)  \\
    e & 10 & 0 & 0 & 392 & 401 & 392 &(3)  \\
    \hline
    f & 1 & 0 & 0 & 222 & 229 & 230 &(3)  \\
    f & 10 & 0 & 0 & 321 & 334 & 350 &(3)  \\
    \enddata
    \tablecomments{(1)~\citet{wolf2017}, (2)~\citet{turbet2018}, (3)~\citet{fauchez2019}}
\end{deluxetable}

\begin{deluxetable*}{lccccp{2.2cm} p{2.8cm} p{2.8cm}} \label{tab:scenarios}
    \tabletypesize{\footnotesize}
    \tablecolumns{5}
    \tablecaption{Scenarios assumed as input for the climate-photochemistry model to simulate the atmosphere of TRAPPIST-1 planets. The relative humidity (RH) is assumed to be constant up to the cold trap. The surface fluxes are the same as for pre-industrial Earth (see Table~\ref{tab:Earth_flux}). For wet \& alive and wet \& dead we assume $\nu_{\text{dep}}$ for O$_2$ and CO according to the underlying scenario. For all other species the $\nu_{\text{dep}}$ shown in Table~\ref{tab:Earth_vdep} are used. 
    For each scenario we assume a range of CO$_2$ surface partial pressures. N$_2$ serves as a fill gas to reach the assumed surface pressure, p$_0$.}
    
    \tablehead{\colhead{Scenario} & \colhead{Planet} & \colhead{CO$_2$ (bar)} & \colhead{$p_0$ (bar)} &  \colhead{RH} & \colhead{Surface flux}  & O$_2$ $\nu_{\text{dep}}$ (cm~s$^{-1}$) & CO $\nu_{\text{dep}}$ (cm~s$^{-1}$)}
    \startdata
        \multirow{ 6}{*}{Wet \& alive}  &  TRAPPIST-1 e & 10$^{-3}$ & 1.001 & \multirow{ 6}{*}{80\%} & \multirow{ 6}{2.2cm}{Biogenic and Volcanic (see~Table~\ref{tab:Earth_flux})} & \multirow{ 6}{2.8cm}{1$\cdot$10$^{-8}$}  & \multirow{ 6}{2.8cm}{3$\cdot$10$^{-2}$ (1$\cdot$10$^{-8}$)} \\
        & TRAPPIST-1 e & 0.01 & 1.01 \\
        & TRAPPIST-1 e & 0.1 & 1.1 \\
        & TRAPPIST-1 e & 1.0 & 2.0 \\
        & TRAPPIST-1 f & 3.6 & 4.0 \\
        & TRAPPIST-1 f & 10.8 & 12.0 \\
        \hline
        \multirow{ 6}{*}{Wet \& dead}  &  TRAPPIST-1 e & 10$^{-3}$ & 1.001 & \multirow{ 6}{*}{80\%} & \multirow{ 6}{2.2cm}{Volcanic (see~Table~\ref{tab:Earth_flux})} & \multirow{ 6}{2.8cm}{1.5$\cdot$10$^{-4}$ (1$\cdot$10$^{-8}$)}  & \multirow{ 6}{2.8cm}{1.2$\cdot$10$^{-4}$ (1$\cdot$10$^{-8}$)} \\
        & TRAPPIST-1 e & 0.01 & 1.01 \\
        & TRAPPIST-1 e & 0.1 & 1.1 \\
        & TRAPPIST-1 e & 1.0 & 2.0 \\
        & TRAPPIST-1 f & 3.6 & 4.0 \\
        & TRAPPIST-1 f & 10.8 & 12.0 \\
        \hline
        \multirow{ 6}{*}{Dry \& dead}  &  TRAPPIST-1 e & 10$^{-3}$ & 1.001 & \multirow{ 6}{*}{80\%} & \multirow{ 6}{2.2cm}{Volcanic (see~Table~\ref{tab:Earth_flux})} & \multirow{ 6}{2.8cm}{1$\cdot$10$^{-8}$}  & \multirow{ 6}{2.8cm}{1$\cdot$10$^{-8}$} \\
        & TRAPPIST-1 e & 0.01 & 1.01\\
        & TRAPPIST-1 e & 0.1 & 1.1\\
        & TRAPPIST-1 e & 1.0 & 2.0 \\
        & TRAPPIST-1 f & 3.6 & 4.0 \\
        & TRAPPIST-1 f & 10.8 & 12.0 \\
    \enddata
        \tablecomments{CO$_2$-poor atmosphere of TRAPPIST-1~e with CO$_2$ partial pressures of only 10$^{-3}$ and 0.01~bar correspond to a $T_{\text{surf}}$ for the wet \& alive run of about 250~K and 260~K, respectively. CO$_2$ partial pressures of 0.1~bar and 3.6~bar for TRAPPIST-1~e and TRAPPIST-1~f, respectively, correspond to a $T_{\text{surf}}$ of about 273~K for the wet \& alive run. CO$_2$ partial pressures of 1~bar and 10.8~bar for TRAPPIST-1~e and TRAPPIST-1~f, respectively, correspond to a $T_{\text{surf}}$ of about 340~K for the wet \& alive run. \\
        O$_2$ deposition is 1$\cdot$10$^{-8}$ for an ocean saturated with O$_2$ (wet \& alive) and for dry \& dead conditions without effective O$_2$ surface sinks \citep{arney2016}. For wet \& dead conditions we assume that the ocean is either saturated or the ocean takes up the O$_2$ with a $\nu_{\text{dep}}$ of 1.5$\cdot$10$^{-4}$~cm~s$^{-1}$ \citep{domagal2014,catling2017}. \citet{schwieterman2019} used a similar value of $\nu_{\text{dep}}$ = 1.4$\cdot$10$^{-4}$~cm~s$^{-1}$ for anoxic atmospheres. \\
        For wet \& alive conditions we assume the same CO deposition of $\nu_{\text{dep}}$ = 3$\cdot$10$^{-2}$~cm~s$^{-1}$ as on Earth \citep{hauglustaine1994,sanhueza1998}, which is larger than the $\nu_{\text{dep}}$ of 1.2$\cdot$10$^{-4}$~cm~s$^{-1}$ calculated for anoxic wet atmospheres \citep{kharecha2005,domagal2014,catling2017,schwieterman2019}. For conditions without effective CO surface sinks we use a $\nu_{\text{dep}}$ of 1$\cdot$10$^{-8}$~cm~s$^{-1}$ \citep{kharecha2005,hu2020}. }
\end{deluxetable*}


\subsection{System parameters and habitability}

We use the following stellar parameters of TRAPPIST-1: a $T_{\text{eff}}$ of 2516~K \citep{vangrootel2018}, a radius of 0.124~$R_{\sun}$ \citep{kane2018}, a mass of 0.089~$M_{\sun}$ \citep{vangrootel2018} and a distance of 12.43~pc \citep{kane2018}. 
Table~\ref{tab:planet_par} provides the planetary parameters for planet e and f used to model the atmosphere and to calculate the S/N of the produced transmission spectra.
We do not focus here on TRAPPIST-1~g since initial studies with our model (not shown) suggested cold, non-habitable conditions, even assuming several tens of bar of surface CO2, although this is a subject for future study \citep[see e.g.][]{wolf2017,turbet2018,lincowski2018}.

Most previous studies used the planetary parameters from \citet{gillon2017} with an irradiation of 0.662~S$_{\sun}$ for TRAPPIST-1~e and an irradiation of 0.382~S$_{\sun}$ for planet~f.
In Table~\ref{tab:tsurf_3d} we compare the mean surface temperature for different atmospheric compositions and using the irradiation from \citet{gillon2017} and \citet{delrez2018b}. We also compare the temperatures with results from 3D studies. 

1D models have difficulties to simulate the atmosphere of planets orbiting low-mass stars in synchronous rotation self-consistently \citep[see e.g.][]{yang2013,leconte2015,barnes2017}. However, Table~\ref{tab:planet_par} shows that the surface temperatures predicted with our 1D model are in general agreement with the results from 3D studies. Using the stellar irradiation from \citet{gillon2017} we overestimate the temperatures by $\sim$10~K for TRAPPIST-1~e. Only for the Titan-like atmosphere with 0.01~bar CH$_4$ and 1~bar N$_2$ do we predict a larger difference of 20~K. For a 10~bar CO$_2$ atmosphere of TRAPPIST-1~f we obtain a 16~K lower surface temperature compared to \citet{fauchez2019}. Note that we only simulate cloud-free conditions. The consideration of clouds in 1D models would likely but not always lead to a lower surface temperature \citep[see e.g.][]{kitzmann2010,lincowski2018}.

\subsection{Model scenarios} \label{ssec:scenarios}

As input for the model we use the SEDs shown in Figure~\ref{fig:T1_spec_new}. 
The atmosphere in the climate module is divided into 101 pressure levels and the chemistry model into 100 altitude layers. We use the full photochemical network with 1127 reactions for 128 species. 

Motivated by the fact that liquid water is a key requirement of life as we know it, we focus here on TRAPPIST-1 e and f, which are found to be favored candidates for habitability \citep[see e.g.][]{wolf2017, turbet2018}. 

We simulate N$_2$ and CO$_2$-dominated atmospheres for TRAPPIST-1~e and CO$_2$-dominated atmospheres for TRAPPIST-1~f. Table~\ref{tab:scenarios} shows the assumed surface pressure, $p_0$, and the surface partial pressure of CO$_2$. N$_2$ serves as a fill gas for each simulation. The partial pressures of CO$_2$ are chosen according to the amount necessary to reach a surface temperature of $\sim$273~K (0.1~bar for planet e and 3.6~bar for planet~f) and $\sim$340~K (1.0~bar for planet e and 10.8~bar for planet~f). According to \citet{wordsworth2013} water loss due to H$_2$O photolysis and hydrogen escape is expected to be weak for surface temperatures below 340~K \citep[see also][]{kasting1993}. 
For TRAPPIST-1~e we additionally use lower CO$_2$ partial pressures
of 10$^{-3}$~bar and 0.01~bar in order to compare with \citet{hu2020} who predicted the composition profiles of TRAPPIST-1~e and f with a 1D photochemistry model using the 3D model output from \citet{wolf2017}.

We assume three scenarios regarding the lower boundary condition: a wet \& alive atmosphere with an ocean as well as biogenic and volcanic fluxes as on Earth, a wet \& dead atmosphere with an ocean and only volcanic outgassing and a dry \& dead atmosphere without an ocean and with only volcanic outgassing (see Table~\ref{tab:scenarios}). We use the same surface pressure for all three scenarios having the same partial pressure of CO$_2$. Hence, depending on the amount of other species in the planetary atmosphere, such as O$_2$ or CO the amount of N$_2$ differs between the scenarios. However, a difference of the surface pressure impedes the comparison between the scenarios due to effects which are not entirely related to the atmospheric composition, such as the surface temperature, pressure broadening, CIA, the eddy diffusion profile and the H$_2$O profile in the lower atmosphere.  

Biogenic and volcanic surface emissions are the same as measured for Earth (see Table~\ref{tab:Earth_flux}). The $\nu_{\text{dep}}$ of CO and O$_2$ are shown in Table~\ref{tab:Earth_vdep}. For all other species we assume a $\nu_{\text{dep}}$ as measured for Earth (see Table~\ref{tab:Earth_vdep}). From \citet{huang2018} we calculate that the net O$_2$ emissions into the atmosphere is 1.29$\cdot$10$^{12}$~molecules~cm$^{-2}$~s$^{-1}$ (11,030~Tg/yr) without taking into account fossil fuel combustion. To reproduce an O$_2$ mixing ratio of 0.21 for our Earth validation run (in Section~\ref{sapp:earth}) we need to set a $\nu_{\text{dep}}$ of 2$\cdot$10$^{-8}$~cm~s$^{-1}$ (not shown) which is similar to the O$_2$ $\nu_{\text{dep}}$=1$\cdot$10$^{-8}$~cm~s$^{-1}$ used by \citet{arney2016}. Hence, we use the value used by \citet{arney2016} as a lower limit for the deposition velocity of O$_2$. The corresponding $\Phi_{\text{BOA}}$ is 1.12$\cdot$10$^{12}$~molecules~cm$^{-2}$~s$^{-1}$ to obtain an O$_2$ mixing ratio of 0.21 with our Earth validation run. 
The escape rates of H, H$_2$ and O are calculated according to the parametrizations presented in Section~\ref{ssec:photo_model}.

\section{Results} \label{sec:results}


\subsection{Atmospheric profiles of TRAPPIST-1~e with 0.1~bar CO$_2$} \label{ssec:atm_T1e}


In this Section we discuss the resulting atmospheric profiles of TRAPPIST-1~e assuming a 0.1 bar surface partial pressure of CO$_2$ in a 1~bar atmosphere. As model input we use all three TRAPPIST-1 spectra from Figure~\ref{fig:T1_spec_new} and compare the resulting atmospheric composition.

\picpage{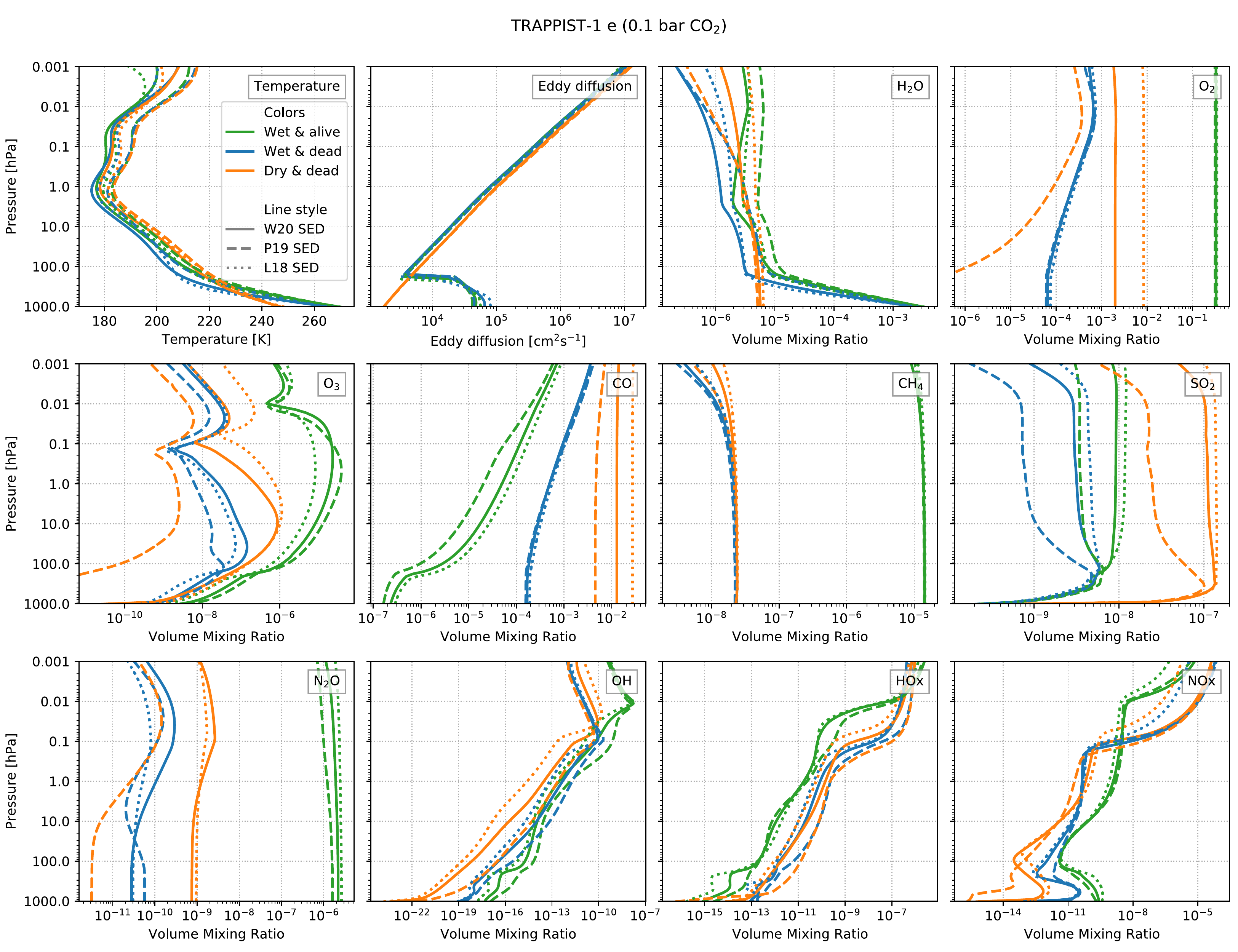}{Temperature, eddy diffusion coefficients and composition profiles of TRAPPIST-1~e runs with 0.1~bar CO$_2$. Different colors represent the three scenarios considered: green for wet \& alive, blue for wet \& dead and orange for dry \& dead. Solid lines represent results using the input TRAPPIST-1 W20 SED, dashed lines show profiles using the P19 SED and dotted lines represent the output using the L18 SED (see also Figure~\ref{fig:T1_spec_new}). }

\subsubsection{Temperature}

\begin{deluxetable}{lccc}\label{tab:tsurf}
    \tabletypesize{\footnotesize}
    \tablecolumns{4}
    \tablecaption{$T_{\text{surf}}$ in K of TRAPPIST-1~e for all three scenarios with 0.1~bar CO$_2$ and different input SED of TRAPPIST-1.}
    \tablehead{\colhead{Input SED}  & \colhead{Wet \& alive} & \colhead{Wet \& dead} & \colhead{Dry \& dead}}  
    \startdata
        W20 SED & 273.1 & 269.6 & 251.5 \\
        P19 SED & 272.2 & 268.2 & 250.4 \\
        L18 SED & 273.7 & 270.9 & 252.7 \\
    \enddata
\end{deluxetable}

Figure~\ref{fig:T1e_01CO2-composition_allspec} shows temperature, eddy diffusion coefficient and composition profiles for selected species for TRAPPIST-1~e with 0.1~bar CO$_2$. The different scenarios are distinguished by color and the different stellar input spectra are denoted by different line styles. The temperature profiles are very similar for all runs except near the surface where the greenhouse effect of H$_2$O leads to larger temperatures for the wet scenarios compared to the dry \& dead runs. The temperature inversion in the middle atmosphere is lacking due to weak UV absorption by O$_3$ (see Section~\ref{ssec:O3_comp}). The wet \& alive runs show the largest $T_{\text{surf}}$ due to warming from biogenic species such as CH$_4$ (see Table~\ref{tab:tsurf}). The impact of the different stellar spectra shown in Figure~\ref{fig:T1_spec_new} on the planetary $T_{\text{surf}}$ is generally small.

\subsubsection{Eddy diffusion coefficients}
For the dry scenario the eddy diffusion coefficient, $K$, near the surface is low and increases continuously towards higher altitudes. This is similar to the $K$ profiles estimated for Venus and Mars \citep[e.g.][]{nair1994,krasnopolsky2012}. The wet scenarios follow a $K$ profile which is similar to Earth with a decrease of $K$ up to the cold trap and an increase above \citep{massie1981}. This profile is also similar to that calculated by \citet{lincowski2018} for the atmosphere of TRAPPIST-1~e, assuming an Earth-like planet covered by an ocean. 

\subsubsection{H$_2$O}
The water profile in the lower atmosphere depends mainly on the fixed relative humidity and the temperature. For the wet scenarios the relative humidity profile is assumed to be constant at 80\% in the lower atmosphere. For the dry runs only the surface H$_2$O is calculated with the relative humidity, otherwise the H$_2$O profile is determined chemically. For pressures below 1~hPa H$_2$O is mainly destroyed photochemically at wavelengths shorter than 200~nm and reformed via HO$_x$-driven (HO$_x$ = H + OH + HO$_2$) oxidation of CH$_4$ into H$_2$O. 
The scenario which includes biogenic fluxes of the Earth as additional lower boundary condition (wet \& alive) leads to significant H$_2$O production via CH$_4$ oxidation \citep[see also][]{segura2005,grenfell2013,rugheimer2015,wunderlich2019}.  

\subsubsection{CH$_4$}
The abundances of CH$_4$ are mainly driven by the surface flux. For the alive scenario we use pre-industrial (biogenic and volcanic) flux measured on Earth (6.31$\cdot$10$^{10}$~cm s$^{-1}$, see Table~\ref{tab:Earth_flux}) and for the dead runs we use only geological sources of CH$_4$ (1.12$\cdot$10$^{8}$~cm s$^{-1}$, see Table~\ref{tab:Earth_flux}). The choice of the SED has no impact on the CH$_4$ abundances in the lower atmosphere. For pressures below 0.1~hPa, where destruction of CH$_4$ is dominated by photolysis, the choice of the SED has only a weak impact on the CH$_4$ concentrations. As found in previous works the CH$_4$ abundances are increased for a planet orbiting an M-dwarf compared to a few ppm on Earth \citep[e.g.][]{segura2005, grenfell2013, grenfell2014, rugheimer2015, wunderlich2019}. This is mainly due to reduced sources of OH via e.g. H$_2$O~+~O($^1$D)~~$\rightarrow$~2~OH, where O($^1$D) comes mainly from O$_3$ photolysis in the UV. Cool stars, such as TRAPPIST-1 are weak UV emitters, favoring a slowing in the OH source reaction and less destruction of CH$_4$ by OH \citep[see e.g.][]{grenfell2013}. 

In \citet{wunderlich2019} we modelled an Earth-like planet with Earth's biofluxes around TRAPPIST-1 and found that the atmosphere would accumulate about 3000~ppm of CH$_4$. The much lower value of around 15~ppm suggested by this study is due to two main reasons. First, for this study we only consider the natural sources of CH$_4$, whereas in \citet{wunderlich2019} we also included anthropogenic sources. CH$_4$ emissions similar to modern Earth would correspond to a very short period of Earth's history whereas pre-industrial emissions of CH$_4$ persisted for a much longer time. Second, we consider a non-zero CH$_4$ deposition velocity of 1.55$\cdot$10$^{-4}$~cm/s, reducing the amount of CH$_4$ accumulated in the atmosphere. We use this measured deposition velocity of CH$_4$ to validate our model against Earth (see Section~\ref{sapp:earth}). With a zero deposition we would overestimate modern Earth amounts of CH$_4$ and hence, we also consider a deposition of CH$_4$ for the TRAPPIST-1 planets.


\subsubsection{O$_2$}
The alive scenario assumes a constant Earth-like O$_2$ flux from photosynthesis rather than a constant mixing ratio at the surface. The resulting mixing ratio for TRAPPIST-1~e with 0.1~bar CO$_2$ is around 35~\%. The increase of O$_2$ compared to Earth is consistent with results of \citet{gebauer2018b}, who found that the required flux to reach a certain O$_2$ concentration is reduced on an Earth-like planet around AD Leo compared to the Earth around the Sun. 
This is due to the lower UV flux of M-dwarfs, compared to solar like stars, resulting in weaker destruction of O$_2$ in an Earth-like planetary atmosphere.
However, for an atmosphere with about 0.35 bar O$_2$ forest ecosystems would be unlikely because the frequency of wildfires is expected to be increased, preventing the build-up of larger concentrations of O$_2$ \citep[see e.g.][]{watson1992,kump2008}. This effect is not considered in the model.

For the dry \& dead runs there is a large spread of O$_2$ abundances ranging from surface concentrations below 1~ppm using the P19 SED to almost 1~\% using the L18 SED. This spectrum has the largest stellar FUV/NUV ratio, which was shown to favor the abiotic build-up of O$_2$ in CO$_2$-rich atmospheres as follows \citep[see e.g.][]{selsis2002,segura2007,tian2014,france2016}: CO$_2$ photolysis below 200~nm leads to CO and atomic oxygen.
Then either atomic oxygen produces O$_2$ (by e.g. O~+~O~+~M~$\rightarrow$~O$_2$~+~M or O~+~OH~+~M$\rightarrow$~O$_2$+~H~+~M) or is recombined with CO via the HO$_x$ catalysed reaction sequence, which results overall in CO$_2$ forming: CO~+~O~$\xrightarrow[]{\text{HO$_x$}}$~CO$_2$ \citep[see e.g.][]{selsis2002,domagal2014,gao2015,meadows2017}. The reduced production of HO$_x$ by H$_2$O destruction in the lower atmosphere for the dry \& dead cases, compared to the wet \& dead runs, leads to more favorable conditions for abiotic O$_2$ build-up. Additionally the deposition of O$_2$ into an unsaturated ocean, as assumed for the wet \& dead cases, is stronger than the deposition onto desiccated surfaces for the dry cases \citep[see][]{kharecha2005,domagal2014}. 

\subsubsection{O$_3$} \label{ssec:O3_comp}
The production of O$_3$ in the middle atmosphere depends on the O$_2$ concentration and the UV radiation in the Schumann-Runge bands and Herzberg continuum (from about 170~nm to 240~nm). The destruction of O$_3$ is mainly driven by absorption in the Hartley (200 nm~-~310~nm), Huggins (310~nm -~400~nm), and Chappuis (400~nm~-~850~nm) bands. HO$_x$ and NO$_x$ destroy O$_3$ via catalytic loss cycles in the middle atmosphere \citep[see e.g.][]{brasseur2006,grenfell2013}. For the scenario with constant O$_2$ flux of 1.21$\cdot$10$^{12}$ molecules cm$^{-2}$ s$^{-1}$, more O$_3$ is produced than for the dead runs, where O$_2$ is only produced abiotically. For the L18 SED with lower UV flux between 170 and 240~nm, the O$_3$ layer is weaker than for the runs using the other stellar spectra. Due to enhanced abundances of O$_2$ compared to Earth, we find that more O$_3$ is produced. 
\citet{omalley2017} suggested a weaker O$_3$ layer as on Earth, assuming an O$_2$ surface partial pressure of 0.21~bar. 

\subsubsection{CO}\label{ssec:CO_prof}
Photolysis of CO$_2$ in the UV produces CO and O. The dry scenario builds up more CO than the wet cases. 
For the alive runs with additional O$_2$ surface sources, the CO recombines more efficiently to CO$_2$ (via CO~+~O~$\xrightarrow[]{\text{HO$_x$}}$~CO$_2$), resulting in lower CO amounts compared to the dead runs. Additionally we assume a net deposition of CO from the atmosphere to the soil-vegetation system, reducing the amount of CO accumulated in the atmosphere \citep[e.g.][]{prather1995,sanhueza1998}. As for O$_2$, the abundances of CO are larger for the dry \& dead runs than the wet \& dead runs mainly due to the assumed strong uptake of CO by the ocean for the wet scenario. 

The CO mixing ratios are comparable to the results of \citet{hu2020}. For an atmosphere consisting of 1~bar N$_2$ and 0.1~bar CO$_2$ they suggest a partial pressure of CO of about 0.05~bar using a weak $\nu_{\text{dep}}$ of 1$\cdot$10$^{-8}$~cm/s and a CO partial pressure of $\sim$1$\cdot10^{-4}$~bar assuming a direct recombination reaction of O$_2$ and CO in the ocean.
The less effective build-up of CO and abiotic O$_2$ due to a strong surface sink gives indirect evidence on the presence of a liquid ocean. 
Hence, under the simulated conditions with strong CO$_2$ photolysis, CO could not only serve as an "antibiosignature" gas as discussed in e.g. \citet{zahnle2008,wang2016,nava2016,meadows2017,catling2018} and \citet{schwieterman2019} but would indirectly suggest the absence of a liquid ocean at the surface.

The largest abundances of CO for the dry scenarios are found using the L18 SED. This is due to the lower abundances of HO$_x$, in particular OH, which reduce the recombination of CO~+~O into CO$_2$. In turn, large amounts of HO$_x$, like for the dry scenario using the P19 SED, lead to low build-up of CO.


\picfull{T1e_01CO2_spectrum_allspec}{Simulated atmospheric features of the TRAPPIST-1~e runs with 0.1~bar CO$_2$, represented by cloud-free transit transmission spectra and binned to a constant resolving power of $R$=300 (maximum resolving power of NIRSpec PRISM at 5~$\mu$m). Important atmospheric molecular absorption bands are highlighted with horizontal lines in the color of the scenario with the strongest feature or in gray when all scenarios show a strong feature.}

\pic{T1e_01CO2_spectrum_haze}{Simulated transmission spectrum of the TRAPPIST-1~e wet \& alive run with 0.1~bar CO$_2$ with and without the impact of aerosol extinction.}

\subsubsection{SO$_2$}
The main source of SO$_2$ is volcanic outgassing, which is assumed to be equally distributed over the first 10~km of the atmosphere. For a 1~bar N$_2$ atmosphere with 0.1~bar CO$_2$, this corresponds to pressure levels below $\sim$250~hPa. The large $\nu_{\text{dep}}$ of 1~cm/s \citep{sehmel1980} leads to a strong decrease of SO$_2$ towards the surface for all three scenarios. Due to its large solubility in water, SO$_2$ is deposited easily over wet surfaces, such as oceans.  However, \citet{nowlan2014} showed that over desert areas the $\nu_{\text{dep}}$ of SO$_2$ is approximately 0.5~cm/s, hence our value of 1~cm/s which is applied for dry cases as well may overestimate the deposition.

For the wet scenarios we assume Earth-like wet deposition following \citet{giorgi1985}. Most SO$_2$ dissolves into condensed water and is rained out of the atmosphere as sulfate. This process greatly decreases the mixing ratio of SO$_2$ for the wet cases but not for the dry scenarios. 

The remaining SO$_2$ is transported upwards and is partly destroyed by photolysis. 
SO$_2$ photodissociates below 400~nm and strongest below 250~nm \citep[e.g.][]{manatt1993}. Hence, for the scenarios using the P19 SED we find the strongest destruction of SO$_2$ above 100~hPa. 


\subsubsection{N$_2$O}
The main N$_2$O source on Earth are surface biomass emissions. For the alive scenario we find concentrations of N$_2$O comparable to previous studies such as \citet{rugheimer2015} and \citet{wunderlich2019}. The photodissociation of N$_2$O is closely related to the SED around 180 nm \citep[e.g.][]{selwyn1977}, leading to lower abundances of N$_2$O using the P19 SED. 


\subsection{Transmission spectra of TRAPPIST-1~e with 0.1~bar~CO$_2$} \label{ssec:transmissions_T1e}

Figure~\ref{fig:T1e_01CO2_spectrum_allspec} shows the simulated transmission spectra of the TRAPPIST-1~e atmosphere scenarios with surface partial pressures of 0.1~bar CO$_2$, binned to a constant resolving power of $R$=300. The spectra are simulated by the GARLIC model taking as input the chemical and temperature profiles discussed in Section~\ref{ssec:atm_T1e}. We do not take into account the effect of clouds but we include weak extinction from aerosols (see Fig.~\ref{fig:T1e_01CO2_spectrum_haze}). 

The CO$_2$ absorption features are similarly strong for all runs. The wet \& alive runs show strong absorption of O$_3$ in the VIS at around 600~nm and in the IR at 9.6~$\mu$m. The alive run with the P19 SED shows the largest O$_3$ features, due to the more pronounced O$_3$ layer in the middle atmosphere compared to the runs using the other SEDs. The spectral features of abiotic production of O$_3$ and O$_2$ for the dead runs are generally much weaker than the biogenic features. This suggests that only the O$_3$ feature at 9.6~$\mu$m could lead to a false positive detection of O$_3$.  

The CH$_4$ feature at 2.3~$\mu$m which is visible for the alive runs overlaps in low resolution with the CO feature which occurs for the dead \& dry runs. The dead runs using the W20 and L18 SEDs show much larger absorption of CO at 2.3~$\mu$m than the wet \& dry runs. For the dead runs with the P19 SED wet and dry conditions are not clearly distinguishable due to the weak build-up of CO in the dry run (see Section~\ref{ssec:CO_prof}). 

Weak H$_2$O absorption in the lower atmosphere of the dry runs result in more pronounced spectral windows between e.g. 1.7 and 1.8~$\mu$m. The H$_2$O features between 5.5 and 7~$\mu$m do not show a large difference for the various scenarios since these are dominated by absorption higher up in the atmosphere, where the H$_2$O concentration is predominantly determined by photochemical processes and similar for all cases.


\subsection{Atmospheres with increasing CO$_2$} \label{ssec:atm_allco2}

\begin{deluxetable}{lcccc}\label{tab:tsurf_allco2}
    \tabletypesize{\footnotesize}
    \tablecolumns{5}
    \tablecaption{$T_{\text{surf}}$ in K of TRAPPIST-1~e or TRAPPIST-1~f for all scenarios and increasing amount for CO$_2$. The W20 SED is used as input for the atmospheric model.}
    \tablehead{\colhead{Planet} & \colhead{CO$_2$ (bar)}  & \colhead{Wet \& alive} & \colhead{Wet \& dead} & \colhead{Dry \& dead}}  
    \startdata
        e & 10$^{-3}$ & 245.6 & 245.9 & 238.3 \\
        e & 0.01 & 256.7 & 253.3 & 242.7 \\
        e & 0.1 & 273.1 & 269.6 & 251.5 \\
        e & 1 & 335.7 & 331.6 & 281.1 \\
        f & 3.6 &  279.6 & 272.7 & 233.5 \\
        f & 10.8 & 330.2 & 327.0 & 258.9 \\
    \enddata
\end{deluxetable}

\picpage{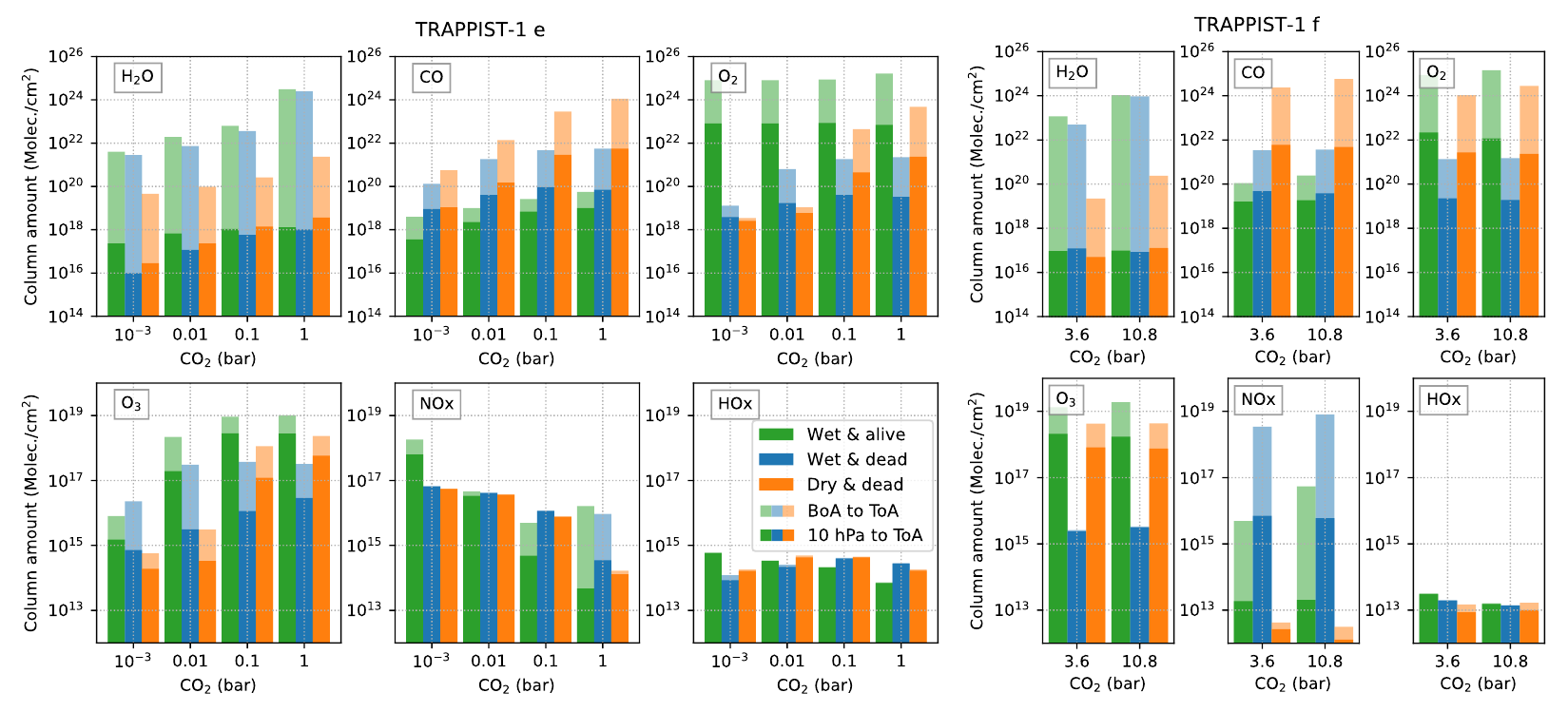}{Column amounts (molecules~cm$^{-2}$) of H$_2$O, CO, O$_2$, O$_3$, HO$_x$ and NO$_x$ for all atmospheric scenarios of TRAPPIST-1~e (left) and TRAPPIST-1~f (right) with increasing partial pressures of CO$_2$ (see also Table~\ref{tab:scenarios}). Semi transparent bars show column amounts integrated from BoA to ToA and  full filled bars show column amounts integrated from 10~hPa to ToA.}

Figure~\ref{fig:T1_column2layer_sem} shows the column amount of H$_2$O, CO, O$_2$, O$_3$, NO$_x$ and HO$_x$ for all three scenarios and with increasing partial pressures of CO$_2$ for TRAPPIST-1~e (left) and TRAPPIST-1~f (right). Semi transparent bars represent column amounts integrated over the entire atmosphere whereas solid filled bars show upper column amounts integrated at pressures below 10~hPa, dominated by photochemical processes. 
For simulations shown in Figure \ref{fig:T1_column2layer_sem} we use the W20 SED as input for the climate-chemistry model.

\subsubsection{H$_2$O}
The H$_2$O amount near the surface mainly depends on the relative humidity and the near surface temperature, leading to an increase of the H$_2$O amount towards larger CO$_2$ partial pressures. Whereas the dry runs show a lower H$_2$O content integrated over the entire atmosphere than the wet runs, at pressures below 10~hPa the three scenarios are comparable (see also Fig. \ref{fig:T1_spec_new}). The $T_{\text{surf}}$ for TRAPPIST-1~e with 1~bar CO$_2$ and TRAPPIST-1~f with 10.8~bar CO$_2$ is $\sim$~340~K for the wet runs. While the total H$_2$O amount increases for an increasing $T_{\text{surf}}$, the increase in the upper atmospheric column is much less, which suggests that tropospheric climate is difficult to elucidate from observing middle atmosphere H$_2$O. Further, the mixing ratio below 10$^{-5}$ (see Fig.~\ref{fig:T1e_01CO2-composition_allspec}) suggests that H$_2$O loss due to H$_2$O photolysis and hydrogen escape is expected to be weak for CO$_2$-rich atmospheres according to \citet{wordsworth2013}. 

\subsubsection{CO}
As discussed in Section~\ref{ssec:atm_T1e} dry \& dead conditions favor an increase in atmospheric CO compared to the wet runs. With increasing CO$_2$ this effect is strengthened due to the enhanced CO$_2$ photolysis for intermediate CO$_2$ amounts. For CO$_2$ partial pressures of 1~bar there is only weak increase of CO column amounts compared to the atmosphere with 0.1~bar CO$_2$, if the $\nu_{\text{dep}}$ of CO is 1$\cdot$10$^{-8}$ cm/s. 
For TRAPPIST-1~f runs with 90\% CO$_2$ there is only weak increase of CO compared to the TRAPPIST-1~e run with 50\% CO$_2$ (1~bar partial pressure of CO$_2$). 
This is consistent with results of \citet{hu2020}. They suggest, that in CO$_2$-rich atmospheres of TRAPPIST-1~e a nonzero deposition velocity of 1$\cdot10^{-8}$cm s$^{-1}$ leads to a maximum build-up of CO of around 0.05~bar.   

\pic{T1_column2layer_sat_sem}{Same as Figure~\ref{fig:T1_column2layer_sem} but with a $\nu_{\text{dep}}$ = 1$\cdot$10$^{-8}$ cm/s for O$_2$ and CO, assuming that the wet runs have an ocean saturated with these gases and the biosphere is not an effective sink for CO. Only O$_2$ and CO are shown because the other species show similar abundances to Figure~\ref{fig:T1_column2layer_sem}.}

For the wet scenarios we assume a much faster deposition of CO due to uptake of the ocean and/or vegetation. The fact that the amount of HO$_x$ is approximately the same for dry and wet surface conditions (see Fig.~\ref{fig:T1_column2layer_sem}), suggests that for wet atmospheres with low CO$_2$ the fast deposition of CO accounts for the weak accumulation of CO. 

We also simulated the abundances of CO and O$_2$ for the wet scenarios assuming that the deposition of CO and O$_2$ into an ocean is weak (see Fig.~\ref{fig:T1_column2layer_sat_sem}). 
We find that the concentrations of CO would be equally high for wet \& dry conditions. Only for the CO$_2$-dominated atmosphere of TRAPPIST-1~f more CO would be present in the dry run compared to the wet runs.

%
\picwide{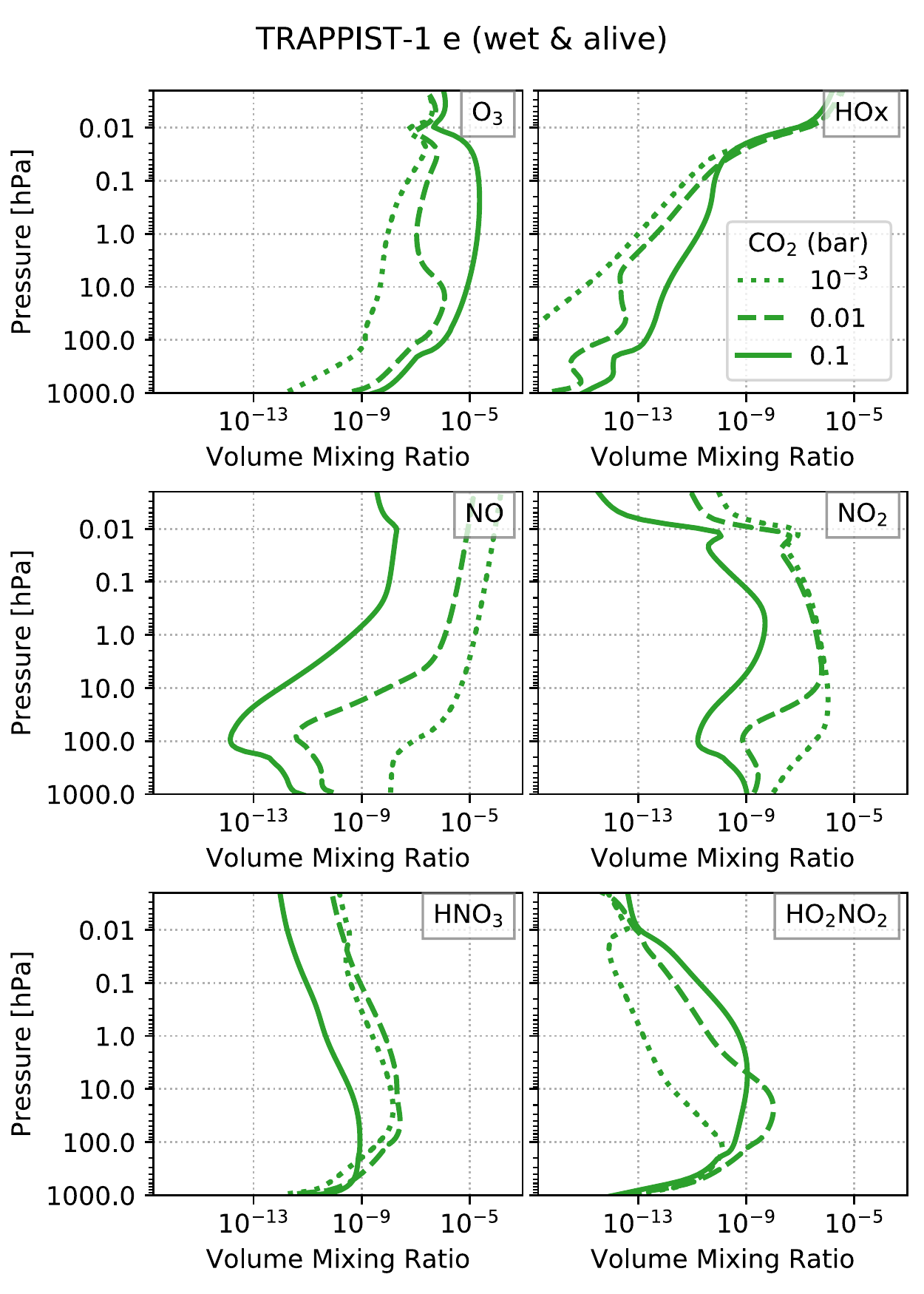}{O$_3$ and related composition profiles of TRAPPIST-1 e wet \& alive runs with 10$^{-3}$~bar CO$_2$ (dotted line), 0.01~bar CO$_2$ (dashed line) and 0.1~bar CO$_2$ (solid line).}

\subsubsection{O$_2$}
For the alive scenario the abundance of O$_2$ is mainly driven by the biogenic surface flux, which is equally strong in all alive runs. Due to the high FUV/NUV ratio for TRAPPIST-1 we expect that significant amounts of O$_2$ are produced abiotically from CO$_2$ photolysis. The potentially false positive detection of O$_2$ in CO$_2$ atmospheres was already discussed by several studies \citep[e.g.][]{selsis2002,segura2007,harman2015,meadows2017}. Figure~\ref{fig:T1e_01CO2-composition_allspec} shows that the abundances of abiotic O$_2$ increase for dry CO$_2$-dominated atmospheres but are always lower than expected from a biosphere similar to the Earth. On the other hand for wet conditions without a biosphere much less abiotic O$_2$ is accumulated in a CO$_2$-dominated atmosphere. This means that weak biogenic O$_2$ flux would not be distinguishable from a dry N$_2$ atmosphere with at least 0.1~bar CO$_2$. 

\subsubsection{O$_3$, NO$_x$ and HO$_x$}
The three scenarios show a different O$_3$ behaviour with increasing CO$_2$ (see Fig.~\ref{fig:T1_column2layer_sem}). The alive run with the lowest amount of CO$_2$ accumulates large amounts of NO$_x$, destroying most of the O$_3$. With increasing abundances of CO$_2$, the temperature increases (see Table~\ref{tab:tsurf_allco2}) and more H$_2$O evaporates. This leads to more HO$_x$ near the surface, more removal of NO$_x$ into reservoirs such as HO$_2$NO$_2$ and less catalytic destruction of O$_3$ by NO$_x$ (see Fig.~\ref{fig:T1e_NOx}).

For the dead runs the O$_3$ is produced abiotically and increases for atmospheres with more CO$_2$. The dry \& dead runs have rather low concentrations of NO$_x$ and HO$_x$ for CO$_2$-dominated atmospheres, which suggests a weak gas-phase effect upon O$_3$ for these species. In contrast, the wet \& dead conditions lead to a build-up of NO$_x$ for TRAPPIST-1~f with 90\% CO$_2$ near the surface, resulting in very low O$_3$ abundances in the lower atmosphere. 

\subsection{Transmission spectra for increasing CO$_2$} \label{ssec:transmission_all}

\picfullwide{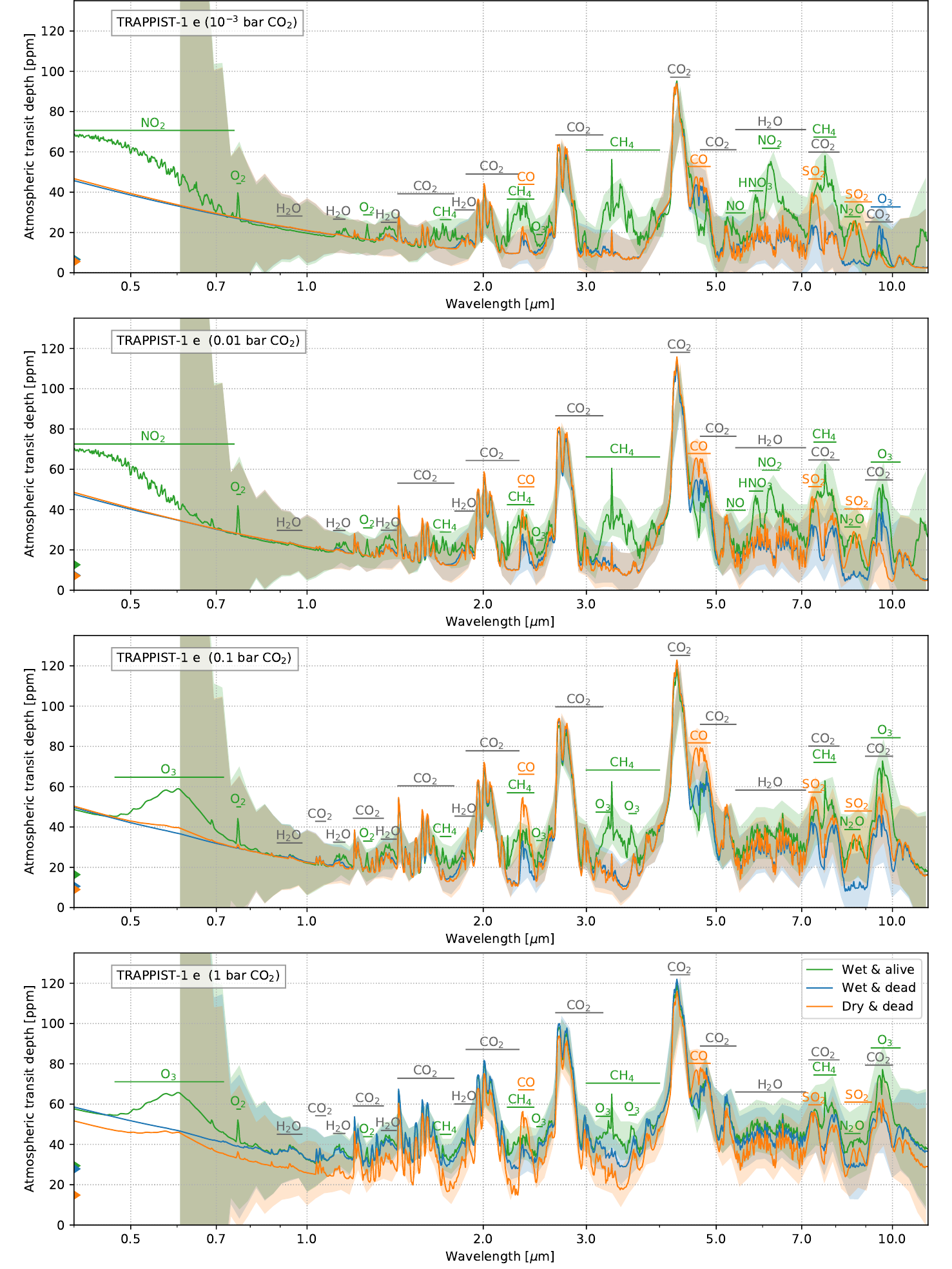}{Simulated atmospheric features of TRAPPIST-1~e runs, represented by cloud-free transit transmission spectra and binned to a constant resolving power of $R$=300 (maximum resolving power of NIRSpec PRISM at 5~$\mu$m). Individual plots from top to bottom show atmospheres with increasing partial pressures of CO$_2$. Shaded regions represent the one sigma error of 30 co-added transit observations with JWST NIRSpec PRISM and MIRI LRS, binned to $R$=30 (minimum resolving power of NIRSpec PRISM at 1~$\mu$m). Important atmospheric molecular absorption bands are highlighted with horizontal lines in the color of the scenario with the strongest feature or in gray when all scenarios show significant features. Coloured triangles indicate minimum atmospheric transit depth of each scenario.}

Figure~\ref{fig:T1e_Spectrum_SNR_lres_full} shows the simulated atmospheric appearance of TRAPPIST-1~e during primary transit for the three scenarios and for increasing amounts of CO$_2$. 
Several molecular features distinguish the alive run with 10$^{-3}$~bar CO$_2$ from the dead runs. Features from CH$_4$, O$_2$ and N$_2$O are present due to the assumed biogenic flux. Strong CH$_4$ features are especially prevalent in Earth-like atmospheres with low UV environments in the habitable zone around M-dwarfs \citep[e.g.][]{segura2005,rauer2011,wunderlich2019}. 
Additionally we find a strong NO$_2$ feature in the VIS as well as NO$_2$, NO and HNO$_3$ features between 5 and 7~$\mu$m due to the large amounts of these species in cold, CO$_2$-poor alive runs (see Fig.~\ref{fig:T1e_NOx}). 
These features are found to be present also for strong flaring conditions with cosmic-ray-induced amounts of NO$_2$ \citep[see e.g.][]{tabataba2016,scheucher2018,scheucher2020a}. 
The typical O$_3$ band around 9.6~$\mu$m is absent due to the large abundances of NO$_x$ species, which can destroy O$_3$ catalytically. 

The dead runs with low CO$_2$ abundances show little spectral differences between wet and dry scenarios. Only SO$_2$ features around 7.5 and 8.5~$\mu$m and weak CO bands around 2.3~$\mu$m and 4.7~$\mu$m distinguish the dry \& dead run from the wet \& dead run. 
With increasing CO$_2$ the larger abundances of CO for dry \& dead conditions lead to stronger CO absorption bands and clearly separate dry from wet runs. 
The presence of the CO bands for CO$_2$-rich atmospheres was also shown by e.g. \citet{meadows2017} and \citet{schwieterman2019}.

For CO$_2$ partial pressures of 0.1~bar and above, NO$_x$ is reduced and its spectral features do not appear in the transmission spectrum. O$_3$ abundances are increased and molecular bands show up in the VIS and at 9.6~$\mu$m. The CH$_4$ abundances are very similar for all runs and hence the CH$_4$ absorption at 2.3~$\mu$m and 3.3~$\mu$m for a 1~bar CO$_2$ should be as strong as for a CO$_2$-poor atmosphere. However, the increase in CO$_2$ abundances lead to larger lower atmosphere temperatures, hence more H$_2$O in this region. Since H$_2$O absorbs over a wide wavelength range this results in an increase in the offset of the entire spectrum \citep[see e.g.][]{turbet2019}, reducing the CH$_4$ features relative to the overall absorption. This is also suggested by Table~\ref{tab:baseline}, showing the baseline of TRAPPIST-1~e transmission spectra from Figure~\ref{fig:T1e_Spectrum_SNR_lres_full}.

\begin{deluxetable}{lcccccc}\label{tab:baseline}
    \tabletypesize{\footnotesize}
    \tablecolumns{7}
    \tablecaption{Minimum atmospheric transit depth, $t_{\text{min}}$ (ppm)  and corresponding $\lambda$ ($\mu$m) of the transmission spectra of TRAPPIST-1~e for all three scenarios and different amount of CO$_2$. $t_{\text{min}}$ is calculated for a constant $R$ of 300 in the NIRSpec PRISM wavelength range (0.6 - 5.3~$\mu$m).}
    \tablehead{\colhead{} & \multicolumn{2}{c}{Wet \& alive}   & \multicolumn{2}{c}{Wet \& dead}  & \multicolumn{2}{c}{Dry \& dead}  \\
    CO$_2$  & $t_{\text{min}}$  & $\lambda$ & $t_{\text{min}}$  & $\lambda$  & $t_{\text{min}}$  & $\lambda$  \\
    (bar) & (ppm) & ($\mu$m) & (ppm) & ($\mu$m) & (ppm) & ($\mu$m)}  
    \startdata
    10$^{-3}$ & 9.44  & 3.06 & 6.51  & 3.51 & 6.39  & 3.51  \\
    0.01      & 12.63 & 2.14 & 7.39  & 3.51 & 7.11  & 3.51  \\
    0.1       & 16.37 & 1.51 & 10.59 & 3.51 & 8.96  & 3.51  \\
    1         & 29.44 & 1.25 & 27.86 & 2.24 & 14.84 & 2.24  \\
    \enddata
        \tablecomments{$t_{\text{min}}$ depends on $R$ and the considered wavelength range. }
\end{deluxetable}

\picwide{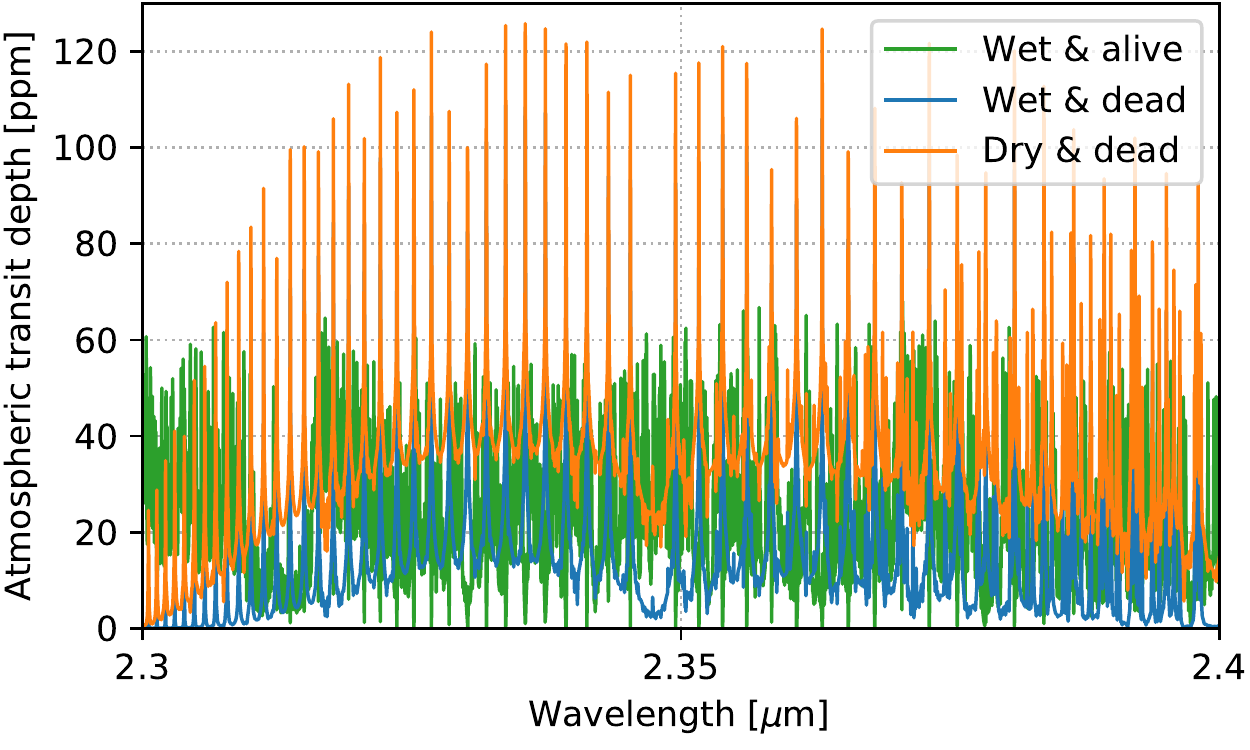}{High resolution transmission spectra of TRAPPIST-1~e runs with 0.1~bar CO$_2$ with a resolving power of $R$=100,000, appropriate for the ELT HIRES. Green line: CH$_4$ features of the wet \& alive run. Blue lines: CO features of the wet \& dead run. Orange line: CO features of the dry \& dead run. Absorption from species other than CO or CH$_4$ are subtracted from the spectrum.}

\picfullwide{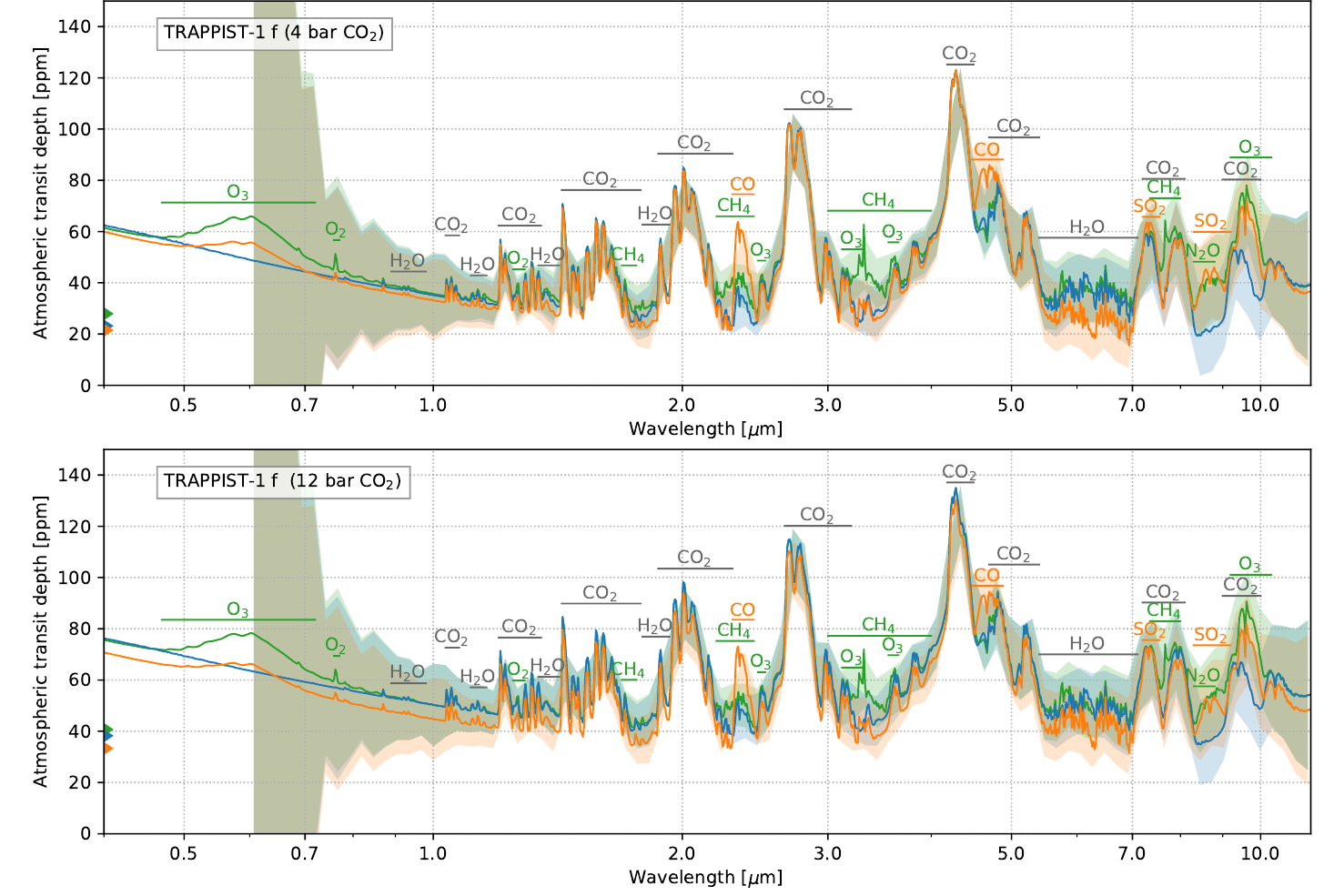}{Same as Figure~\ref{fig:T1e_Spectrum_SNR_lres_full} but for TRAPPIST-1~f runs.}

The most promising candidates for distinguishing the three scenarios from each other are the CH$_4$ features, which are just evident in the alive runs as well as strong CO bands for the dry runs. Absorption of CH$_4$ and CO features overlap at 2.3~-~2.5~$\mu$m, which could inhibit their separation. A simultaneous observation of CH$_4$ at 3.3~$\mu$m is therefore required as well as measurements of CO at 4.6~$\mu$m. Using JWST NIRSpec PRISM covers 0.60 - 5.30 $\mu$m, however TRAPPIST-1 is close to the saturation limit of NIRSpec PRISM (J $<$ 10.5), resulting in a low duty cycle \citep[see e.g.][]{batalha2017}. We do not consider a partial saturation strategy to improve the duty cycle as suggested by \citet{batalha2018}. NIRSpec G235M only covers 1.66 - 3.12 $\mu$m, hence would not be suitable for separating CH$_4$ and CO. Another possibility to disentangle both features is by observing individual lines with high resolution spectroscopy (HRS). 
Figure~\ref{fig:T1e_01CO2_Spectrum_SNR_hres_CH4_CO} shows the simulated transmission spectra of the TRAPPIST-1~e runs with 0.1~bar CO$_2$, binned to the resolution of ELT HIRES ($R$=100,000). Since the position of the lines relative to each other differ between CO and CH$_4$ one could use the cross-correlation technique to determine which absorber causes the spectral lines or even if both species are present.

The transmission spectra of the TRAPPIST-1~f atmospheres show similar spectral features to those of TRAPPIST-1~e with 1~bar CO$_2$ (see Figure~\ref{fig:T1f_Spectrum_SNR_lres_full}).

\subsection{Detectability of spectral features} \label{ssec:detection}

\begin{deluxetable}{m{1.5cm} m{2.3cm} m{1.05cm} m{0.95cm} m{1.05cm}} \label{tab:Ntr_01CO2}
    \tabletypesize{\footnotesize}
    \tablecolumns{11}
    \tablecaption{Number of transits required to detect spectral features with an S/N of 5 in a cloud-free TRAPPIST-1~e atmosphere with 0.1~bar CO$_2$ using LRS with JWST NIRSpec or JWST MIRI and HRS with ELT HIRES. For LRS, $\lambda$ corresponds to the central wavelength of the spectral feature whereas for HRS the considered wavelength range is given. For JWST NIRSpec the filter with the largest S/N for the spectral feature is considered (see Table~\ref{tab:instruments} and Figure~\ref{fig:SNR_T1_1h_JWST_ELT}). For potentially detectable features the required number of transits  using JWST NIRSpec PRISM is given in parenthesis. Numbers below 30 are highlighted in bold face.}
    \tablehead{\colhead{Telescope} 	&	\colhead{Specie ($\lambda$)}	&	Wet \& alive &	Wet \&	dead &	Dry \& dead}  
    \startdata
    JWST 	&	CO$_2$ (4.3 $\mu$m)	&	\textbf{5 (11)}	&	\textbf{4 (9)}	&	\textbf{4 (8)}	\\
    ELT	&	CO$_2$ (1.8-2.3 $\mu$m)	&	33	&	\textbf{28}	&	\textbf{26}	\\
    \hline
    JWST	&	H$_2$O (1.4 $\mu$m)	&	170	&	107	&	100	\\
    ELT 	&	H$_2$O (1.3-2.0 $\mu$m)	&	1224	&	1424	&	865	\\
    \hline
    JWST	&	CH$_4$ (3.3 $\mu$m)	&	60 (60)	&	-	&	-	\\
    ELT 	&	CH$_4$ (2.1-2.5 $\mu$m)	&	\textbf{26}	&	7,434	&	$>$10,000	\\
    \hline
    JWST 	&	CO (2.35 $\mu$m)	&	-	&	114	&	\textbf{19} (57)	\\
    ELT 	&	CO (2.3-2.45 $\mu$m)	   &	437	&	105	&	42	\\
    \hline
    JWST 	&	O$_3$ (9.6 $\mu$m)	&	124	&	255	&	258	\\
    ELT 	&	O$_3$ (3.4-3.7 $\mu$m)	&	4,024	&   $>$10,000	&	$>$10,000	\\
    \hline
    JWST 	&	O$_2$ (1.27 $\mu$m)	&	3,012	&	-	&	-	\\
    ELT 	&	O$_2$ (1.24-1.3 $\mu$m)	&	910	&	$>$10,000	&	$>$10,000	\\
    \hline
    JWST 	&	SO$_2$ (7.35 $\mu$m)	&	-	&	-	&	146	\\
    ELT 	&	SO$_2$ (3.9-4.1 $\mu$m)	&	-	&	-	&	$>$10,000	\\
    \hline
    JWST 	&	N$_2$O (8.5 $\mu$m)	&	1,292	&	-	&	-	\\
    ELT 	&	N$_2$O (2.1-2.3 $\mu$m)	&	951	&	-	&	-	\\
    \enddata
    \tablecomments{The ETC for the ELT does not include the wavelength range 2.9-3.4~$\mu$m which will be covered by METIS \citep{brandl2016}. Since O$_3$ absorbs in the L-band we might overestimate the number of transits required to detect O$_3$.
}
\end{deluxetable}

\picwide{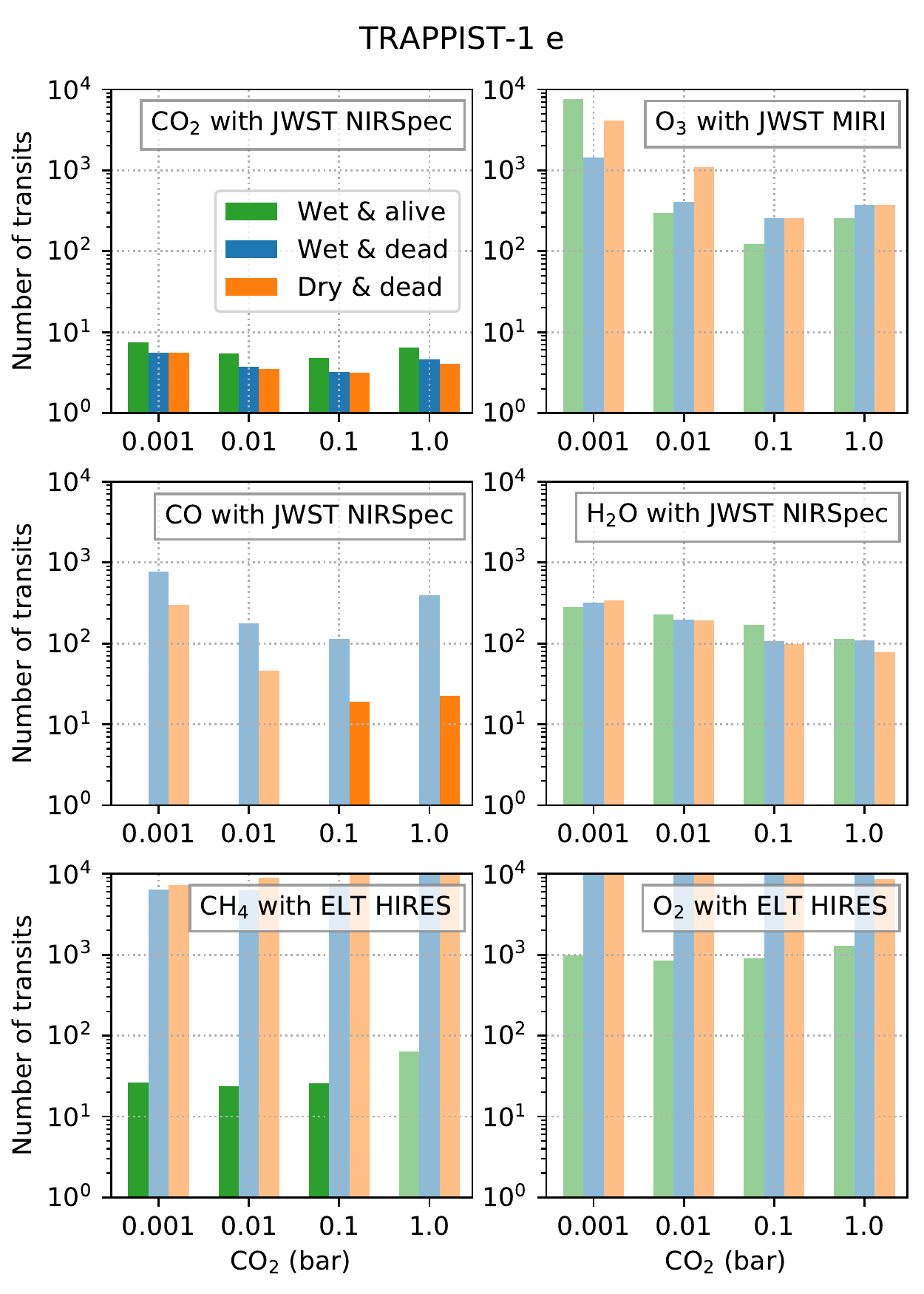}{Number of transits required to reach an S/N of 5 for the corresponding spectral features of CO$_2$ at 4.3~$\mu$m, O$_3$ at 9.6~$\mu$m, CO at 2.35~$\mu$m and H$_2$O at 1.4 $\mu$m with JWST NIRSpec (upper and middle panel) and CH$_4$ from 2.1 to 2.5 $\mu$m and O$_2$ from 1.24 to 1.3 $\mu$m with ELT HIRES (lower panel) in a cloud-free atmosphere of TRAPPIST-1~e. The x-axes correspond to the increasing partial pressures of CO$_2$.
Full filled bars: required number of transits is below or equal 30. Semi transparent bars: required number of transits is larger than 30.
}

\picwide{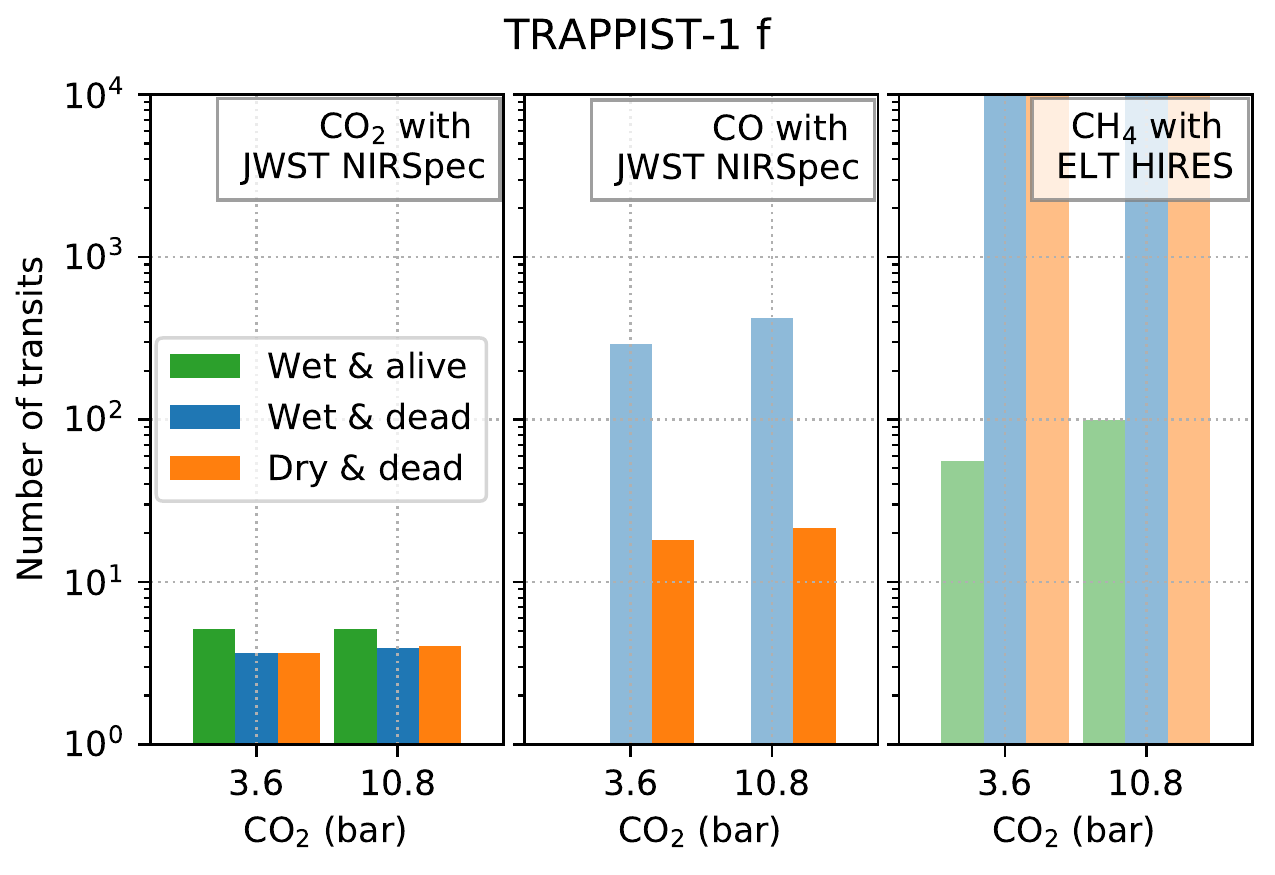}{Number of transits required to reach an S/N of 5 for the corresponding spectral features of CO$_2$ at 4.3~$\mu$m and CO at 2.35~$\mu$m with JWST NIRSpec (left and center) and CH$_4$ from 2.1 to 2.5 $\mu$m with ELT HIRES (right) in a cloud-free atmosphere of TRAPPIST-1~f. The x-axes correspond to the increasing partial pressures of CO$_2$. Full filled bars: required number of transits is below or equal 30. Semi transparent bars: required number of transits is larger than 30.}

We determine the required number of transits necessary to detect a spectral feature (S/N = 5) with JWST NIRSpec or JWST MIRI. We bin the spectral data until the optimal value is found, leading to the lowest required number of transits. Binning the data decreases the noise contamination but if the binned wavelength range is too large, molecular absorption bands and atmospheric windows overlap, leading to a cancellation of the spectral feature. 
Due to the unknown systematic error when binning the synthetic spectral data we assume only white noise. This gives an optimistic estimation on the detection feasibility of the JWST.
Additionally we estimate the number of transits required to detect spectral absorption lines with ELT HIRES using the cross correlation technique without binning the spectral data (see Section \ref{ssec:snr}). 

\subsubsection{CO$_2$}
Table~\ref{tab:Ntr_01CO2} shows the number of transits needed to detect selected spectral features for all three atmospheric scenarios of TRAPPIST-1~e with 0.1~bar CO$_2$. For all the calculations we assume cloud-free atmospheric conditions with weak extinction from aerosols (see Eq.~(\ref{eq:sigma_aer})). 

With JWST NIRSpec G395M/F290LP only about 5 transits are needed to detect the 4.3~$\mu$m CO$_2$ feature in a cloud-free atmosphere. About twice as many transits are required to detect CO$_2$ with NIRSpec PRISM.
This result is in agreement with other studies such as \citet{fauchez2019}, who showed that the CO$_2$ at 4.3~$\mu$m of a 1~bar CO$_2$ atmosphere of TRAPPIST-1~e would be detectable with JWST NIRSpec PRISM by co-adding 9 transits without the existence of clouds. When taking clouds into account, they suggested that 19 transits are required to detect CO$_2$. 
For a ground-based telescope such as ELT at wavelengths longer than 4~$\mu$m, the noise contribution from the Earth's atmosphere leads to very low S/N. The 2.7~$\mu$m feature of CO$_2$ is not observable with ELT. Hence, only the CO$_2$ feature around 2.0~$\mu$m might be detectable with ELT HIRES in $\sim$30 transits. 

The molecular bands for CO$_2$ do not greatly increase when increasing the abundances of CO$_2$ from 10$^{-3}$~bar to 1~bar, hence also the number of transits needed to reach the same S/N of 5 are similar for all runs (see Fig.~\ref{fig:T1e_ntransit_allruns}). It was shown by \citet{barstow2016} that even the Earth and a 1~bar Venus-like atmosphere would show similar CO$_2$ features, which complicates the determination of the underlying atmospheric main composition by retrieval methods. \\ \  \\
%

\subsubsection{H$_2$O}
A larger CO$_2$ partial pressure warms the lower atmosphere, leading to more H$_2$O evaporation in the case of a liquid reservoir. This leads to a more opaque lower atmosphere and an increase in the measured planetary radius \citep[see e.g.][]{vonparis2011,madhusudhan2015}. In contrast, in the photochemical regime, H$_2$O is not greatly increased for warmer surface conditions (see Fig.~\ref{fig:T1_column2layer_sem}). The effect of the radius increase is much weaker for dry atmospheres, leading to a better detectability of H$_2$O for dry surface conditions. However, the H$_2$O spectral features are too weak in all simulated atmospheres of TRAPPIST-1~e and TRAPPIST-1~f to allow for a detection with JWST NIRSpec. This was also concluded by \citet{fauchez2019} who found that about 150 transits are required to detect H$_2$O in a cloud-free 1~bar CO$_2$ atmosphere of TRAPPIST-1~e with JWST. 

Most H$_2$O bands in the NIR overlap with CH$_4$ absorption features. This could cause a false positive detection of H$_2$O for large abundances of CH$_4$ \citep[see e.g.][]{wunderlich2019}. The cross-correlation technique could disentangle H$_2$O from CH$_4$ but we find that by using the largest $\sim$500 H$_2$O lines about $\sim$1000 transits would be needed to detect H$_2$O with ELT HIRES. 

\subsubsection{CH$_4$}
In low CO$_2$ atmospheres with biogenic surface fluxes the number of CH$_4$ lines which we identify is much larger than the H$_2$O lines, enabling a detection of CH$_4$ with less than 30 transits using ELT HIRES. The detection of the simulated levels of CH$_4$ would be challenging with JWST NIRSpec. 

In contrast to the alive runs, no CH$_4$ feature is detectable for the dead runs with only geological sources of CH$_4$. However, since the ability to detect CH$_4$ mainly depends on the assumed surface flux, which could be weaker for a potential biosphere on an M-dwarf planet \citep[e.g.][]{cui2017} or stronger for enhanced volcanic outgassing of CH$_4$, the detection or non-detection of CH$_4$ alone would not confirm or rule-out the existence of a biosphere \citep[see also][]{krissansen2018a}.

\subsubsection{CO}
About 10\% of CO$_2$ are needed to produce enough CO photochemically to enable a detection of its molecular absorption feature at 2.35~$\mu$m in a cloud-free atmosphere with JWST NIRSpec G235M if surface sinks of CO are inefficient. For the wet scenarios, with significant CO uptake by an ocean or a biosphere, results suggest, that CO would not be detectable, even for a CO$_2$-dominated atmosphere. 
The CO feature at 4.6~$\mu$m overlaps with the CO$_2$ absorption, requiring a retrieval analysis to disentangle both signals. Only about 10 transits are needed to detect the 4.6~$\mu$m band with JWST. The G395M filter of JWST would be favorable because the CO$_2$ band at 4.3~$\mu$m and the CO feature at 4.6~$\mu$m could be observed simultaneous.\\
The CO feature at 2.3~$\mu$m does not overlap with other strong absorption features in the transmission spectrum of the dry scenarios.
However, 19 transits are required to detect the CO feature at 2.3~$\mu$m (see Table~\ref{tab:Ntr_01CO2}), twice as many as for the detection of the 4.6~$\mu$m CO feature.\\
The detection of CO with the cross correlation technique has been shown to be feasible for gas giants exoplanets \citep[see e.g.][]{dekok2013,brogi2014}. We find that the detection of CO would require about 40 transits with ELT HIRES in a dry, CO$_2$-rich, cloud-free atmosphere of TRAPPIST-1~e and f.

\subsubsection{Other molecules}
Results suggest, that no other molecular absorption features would be observable with JWST or ELT for the atmospheres considered here. Even a detection of the strong NO$_2$ feature around 6.2~$\mu$m in an alive CO$_2$-poor atmosphere (see green line in top panel of Fig.~\ref{fig:T1e_Spectrum_SNR_lres_full}) would require around 50 transits with JWST MIRI (not shown). The O$_3$, SO$_2$ and N$_2$O features lie in a spectral region where the stellar flux is too low to allow high S/N. The O$_2$ feature is not strong enough for a detection with JWST NIRSpec. As also suggested by \citet{rodler2014} we find that the 1.27~$\mu$m band is more favorable than the 0.76~$\mu$m band for detecting O$_2$ in a planetary atmosphere around a very late M-dwarf. We find that with ELT over 900 transits are required to detect O$_2$ by cross-correlating the lines between 1.24 and 1.3~$\mu$m, assuming an average throughput of 10\% for ELT HIRES.
This is consistent with the results of \citet{rodler2014}, who suggested that hundreds of transits are needed to detect O$_2$ in the atmosphere of Earth around an M7 star at a distance similar to TRAPPIST-1 with ELT using a high resolution spectrograph with a throughput of $\sim$20\% \citep[see][]{origlia2010}.

\picwide{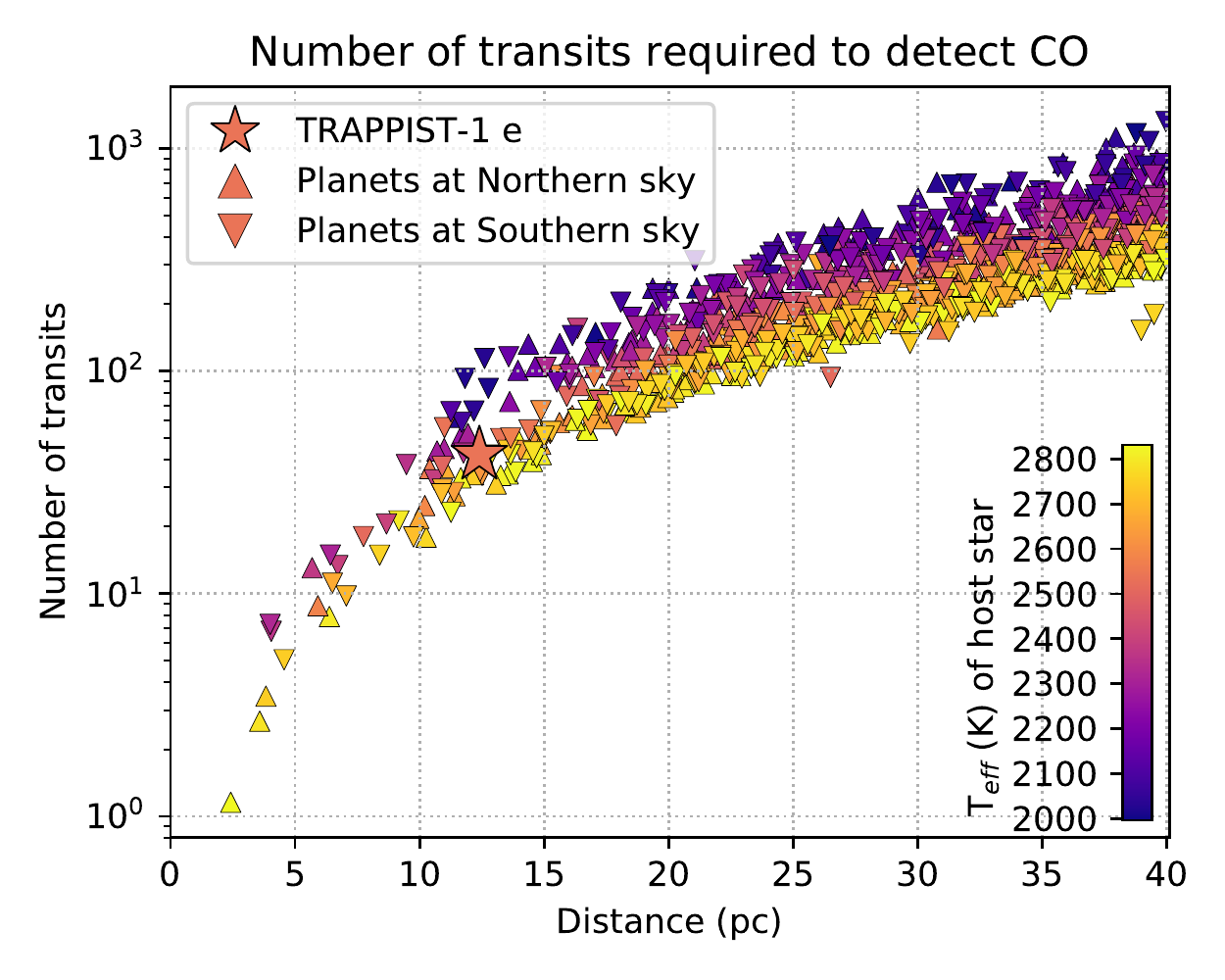}{Number of transits required to detect CO with the cross correlation technique between 2.3~-~2.45~$\mu$m with ELT (Southern sky) or TMT (Northern sky) in the atmosphere of hypothetical planets with the same properties as TRAPPIST-1~e but around SPECULOOS targets. We assume that the atmosphere of all planets is that of the 0.1~CO$_2$ run with dry \& dead conditions. The considered planetary atmospheric spectrum assumes cloud-free conditions. }

\subsubsection{SPECULOOS targets} \label{ssec:speculoos}

With a distance of only 12.4~pc from the Sun, TRAPPIST-1 is one of the closest late-type M-dwarfs. However, we show that for the simulated atmospheres, only CO$_2$ would be potentially detectable within $\sim$10 transits. 
To further characterize the atmosphere of the planets observing the K-band with HRS might allow to determine whether a spectral feature around 2.3~$\mu$m can be attributed to absorption from CH$_4$ or CO. Our results suggest, that for a dry \& dead atmosphere of TRAPPIST-1~e about 40 transits are required to detect CO with ELT HIRES.
To detect an Earth-like O$_2$ feature with the same number of transits, a host star similar to TRAPPIST-1 is required at $\sim$7~pc or less \citep[see][]{rodler2014,serindag2019}.

The Search for habitable Planets EClipsing ULtracOOl Stars  \citep[SPECULOOS;][]{delrez2018} is a ground-based transit survey which is looking for Earth-sized exoplanets around the nearest late M-dwarfs to brown dwarfs. Figure~\ref{fig:Speculoos_snr} shows the number of transits required to detect CO with the cross-correlation technique using TMT (Northern Sky) or ELT (Southern sky), assuming a hypothetical planet with the same properties as TRAPPIST-1~e around each member of the target list of SPECULOOS with a $T_{\text{eff}}$ of at least 2000~K \citep[see][]{gillon2020}. The assumed atmospheric spectral feature is the same as in the dry scenario with 0.1~bar CO$_2$. 

There are only 13 stars within a distance of 7~pc ($T_{\text{eff}}>$2000~K), where the atmospheric O$_2$ feature of a hypothetical terrestrial planet would be detectable within 40 transits according to \citet{rodler2014}. However, non-LTE effects in the O$_2$ 1.27~$\mu$m band may prevent a detection \citep{lopez-puertas2018}. Figure~\ref{fig:Speculoos_snr} suggests that more targets exists for which the CO feature could be detected. For late M-dwarfs (2400~K$~<~T_{\text{eff}}~<$~2800~K) CO could be detected up to $\sim$12pc using ELT or TMT by co-adding 30 transits. Early L-dwarfs ($T_{\text{eff}}<$2400~K) only have slightly smaller stellar radius than late M-dwarfs but are much fainter, resulting in a low S/N and more transits are required to detect atmospheric molecular features with transmission spectroscopy. 

\section{Discussion} \label{sec:discussion}

With our climate-photochemistry model, 1D-TERRA, we simulated potential atmospheres of TRAPPIST-1~e and TRAPPIST-1~f. We determined the composition of the planetary atmospheres, assuming N$_2$ and CO$_2$-dominated atmospheres with wet and dry surface conditions. We did not consider O$_2$-rich atmospheres, accumulated from H$_2$O photolysis during the pre-main sequence phase of TRAPPIST-1 \citep[see e.g.][]{wordsworth2014, luger2015,bolmont2017}. However, an Earth-like biogenic flux of O$_2$ is considered and O$_2$ can also build up abiotically via CO$_2$ photolysis. For detailed discussion of the potential composition and transmission spectra of O$_2$-dominated atmospheres from H$_2$O photolysis we refer to \citet{lincowski2018}. 

The main goal of our study was to investigate which spectral features of wet or dry planets in the habitable zone could be detectable with the upcoming JWST and ELT. We identify three species which could be detectable in a cloud-free atmosphere of TRAPPIST-1 e or f by co-adding less than 30 transits: CO$_2$, CH$_4$ and CO. Under the assumed boundary conditions, CO$_2$ would be detectable with JWST and ELT with about 10 transits. This is also consistent with several other studies investigating the detectability of the atmospheric features of the TRAPPIST-1 planets \citep{morley2017,batalha2018, krissansen2018, wunderlich2019, lustig-yaeger2019, fauchez2019}. 
However, for N$_2$-dominated atmospheres the uncertainties of the retrieved CO$_2$ abundances are up to 2 orders of magnitude when co-adding 10 transits \citep[see][]{batalha2018,krissansen2018}. 

The effect of clouds and hazes is not considered in the model and we only consider weak extinction by aerosols for the simulation of the transmission spectra. We do not expect a large impact on the chemical composition when considering thin cloud or haze layers (see Venus validation, Figure~\ref{fig:Venus}). However, the presence of clouds can significantly reduce the detectability of molecular spectral features \citep[see e.g.][]{kitzmann2011a,kitzmann2011b,vasquez2013,benneke2013,betremieux2014,betremieux2017,moran2018,lustig-yaeger2019,fauchez2019,komacek2020,suissa2020}. 
We use a similar expression to simulate the effect of aerosol absorption to \citet{kaltenegger2009}. They conclude that the apparent radius of an atmosphere like on Earth is mainly determined by Rayleigh scattering and aerosol, H$_2$O and CO$_2$ absorption. For Earth, the inclusion of realistic cloud coverage has only a small effect on the apparent radius and hence, the detectability of spectral features.

For the wet scenarios with low CO$_2$ abundances and Earth-like biomass surface emissions we find that CH$_4$ would be detectable on TRAPPIST-1~e using the cross-correlation technique with less than 30 transits. Increasing the amount of CO$_2$ leads to additional greenhouse warming and more H$_2$O evaporated into the atmosphere. More H$_2$O in the lower atmosphere leads to an increase of the minimum transit depth in the transit spectrum, i.e. the observational baseline \citep[see also][]{turbet2019}. The strongest CH$_4$ feature at 3.3~$\mu$m is about 40~ppm above the baseline, when very little H$_2$O is present in the atmosphere. For a lower atmosphere with a relative humidity of 80\% and a $T_{\text{surf}}$ of $\sim$335~K the baseline increases by 20~ppm compared to a cold atmosphere with a $T_{\text{surf}}$ of $\sim$245~K. Due to this effect, for CO$_2$-dominated atmospheres of TRAPPIST-1~e and TRAPPIST-1~f CH$_4$ would not be detectable for a pre-industrial Earth-like emission flux of CH$_4$, since this feature would be partially swamped by the baseline. For these cases the spectral appearance would not suggest the existence of a biosphere within the detection limits, i.e. it would be a false negative detection of CH$_4$.

Enhanced outgassing when assuming e.g. a more reducing mantle than modern Earth would need to be 2-3 orders of magnitudes larger than for modern Earth to build up as much CH$_4$ as for the alive scenarios \citep[see also][]{ryan2006,krissansen2018a}.
Since also the outgassing of CO is expected to be large for a highly reduced mantle, simultaneous detection of CO could distinguish an atmosphere with large amounts of outgassed abiotic CH$_4$ from an atmosphere with mainly biogenic CH$_4$ \citep[see also][]{krissansen2018a}. 

The presence of large amounts of CO has been suggested to indicate the absence of life on an exoplanet \citep{zahnle2008, wang2016, nava2016, meadows2017, catling2018}. 
We find that the CO feature at 2.3~$\mu$m would be detectable with JWST NIRSpec for a dry atmosphere with at least 0.1~bar CO$_2$ by co-adding $\sim$20 transits (Fig.~\ref{fig:T1e_ntransit_allruns}). In contrast to CH$_4$, CO would be detectable also for CO$_2$-dominated atmospheres due to the enhanced CO build-up from CO$_2$ photolysis. 

The detection of CO with ELT HIRES requires twice as much transits than with JWST when assuming an average throughput of 10\%. Previous studies such as \citet{snellen2013} or \citet{serindag2019} assume a mean throughput of 20\% for ELT. However, to achieve this large efficiency further development of the instrument design might be necessary \citep[see e.g.][]{ben-ami2018}. 

For dry surface conditions, without a liquid ocean, we expect that very little CO would be deposited onto the surface. In contrast, the existence of an ocean may inhibit the build-up of substantial amounts of CO in a CO$_2$-rich atmosphere through catalytic cycles and an effective CO surface sink. This would lead to a non-detection of CO for wet surface conditions.
However, the detection of CO in a CO$_2$-rich atmosphere of an M-dwarf planet could be also compatible with the presence of an ocean and a biosphere with ineffective surface sinks of CO or increased CO surface flux \citep{krissansen2018a,schwieterman2019}. Hence, the detection of CO does not ultimately discriminate between wet and dry surface conditions but a non-detection of CO and a simultaneous detection of CO$_2$ in the atmosphere of a potential habitable TRAPPIST-1 planet can hint at an effective surface sink for CO, suggesting the existence of an ocean. 


As for CO, we find that abundances of SO$_2$ are much larger for dry surface conditions than for wet conditions. For the wet scenarios, most of the SO$_2$ is oxidised into highly soluble sulfate hence efficiently removed from the atmosphere by wet and dry deposition. For the dry scenarios we do not consider any wet deposition. \citet{loftus2019} suggests that the detection of an H$_2$SO$_4$-H$_2$O haze layer together with SO$_2$ indicate that the planet does not host significant surface liquid water. The large amounts of SO$_2$ we find for the dry surface conditions are consistent with their study. However, the detection of SO$_2$ would not be feasible for any of the dry runs of TRAPPIST-1~e and TRAPPIST-1f with JWST or ELT. Furthermore, the SO$_2$ may form a haze layer.

For the simulated N$_2$ and CO$_2$-dominated atmospheres, one would require large observational times to detect spectral features in the atmospheres of the TRAPPIST-1 planets with JWST or ELT \citep[see also][]{morley2017,batalha2018, krissansen2018, wunderlich2019, lustig-yaeger2019,gillon2020}.

In this study we assume white noise only when co-adding multiple transits or binning spectral data to a lower resolution than observed. This assumption may underestimate the required number of transits significantly, especially for weak spectral features \citep[see e.g.][]{fauchez2019}. 
Imaging spectroscopy concepts such as the Large UV/Optical/Infrared Surveyor \citep[LUVOIR, ][]{luvoir2019} and the Habitable Exoplanet Observatory \citep[HabEx, ][]{mennesson2016} may provide new opportunities to observe the atmosphere of terrestrial planets \citep[see e.g.][]{pidhorodetska2020}. The angular separation between TRAPPIST-1 and TRAPPIST-1~e is only 2.4 milliarcseconds (mas), much smaller than for Proxima Centauri b (37 mas) \citep{omalley2019}. This might be too small to separate the star and the planets with LUVOIR or HabEx \citep[see also][]{stark2015}. Hence, transmission spectroscopy is the most promising way to constrain the atmospheric characteristics of the habitable TRAPPIST-1 planets in the next few decades.

The recent detection of H$_2$O absorption in the atmosphere of the habitable zone planet K2-18b is one example of how the existence of an H$_2$ envelope could enable the characterization of the atmosphere of potentially rocky planets \citep{benneke2019,tsiaras2019}. 
Initial observations of the TRAPPIST-1 planets showed no hint of cloud-free H$_2$ or helium dominated atmospheres, suggesting that atmospheres are dominated by heavier elements \citep{dewit2016,dewit2018,wakeford2018,burdanov2019}. However, hydrogen-rich atmospheres with high-altitude clouds or hazes are also consistent with the observations of the TRAPPIST-1 planets \citep{moran2018}. Such hydrogen-rich atmospheres of the planets would increase the scale height, leading to improved detectability of spectral features. 

\section{Summary and Conclusion} \label{sec:conclusion}

We introduced and validated our new chemical network, part of our updated 1D coupled climate-photochemistry model (1D-TERRA). The model is capable of simulating the atmosphere of terrestrial planets over a wide range of temperatures and pressures. Our chemical network is based on those presented by \citet{hu2012} and \citet{arney2016}. Additionally we added chlorine chemistry and extended the sulphur chemistry with chemical reactions listed in \citet{zhang2012}, in order to simulate Venus-like atmospheres. We showed that the model is able to reproduce modern Earth as well as CO$_2$-dominated atmospheres such as present on modern Mars and Venus. The resulting composition profiles are consistent with observations and other photochemical models, dedicated to model the atmosphere of Mars \citep{nair1994, krasnopolsky2010} and Venus \citep{krasnopolsky2012,zhang2012}. 

In this paper we simulated the potential atmospheres of the TRAPPIST-1~e and TRAPPIST-1~f planets assuming N$_2$ and CO$_2$-dominated atmospheres for three main scenarios regarding the lower boundary condition: first, a wet \& alive atmosphere with an ocean as well as biogenic and volcanic fluxes as on Earth, second, a wet \& dead atmosphere with an ocean and only volcanic outgassing and, third, a dry \& dead atmosphere without an ocean and with only volcanic outgassing (see Table~\ref{tab:scenarios}).  
We showed the simulated atmospheric composition and spectral appearance of TRAPPIST-1~e with 0.1~bar CO$_2$ using three different SEDs as input for the climate-chemistry model. To our knowledge ours is the first study which uses an SED of TRAPPIST-1 which was constructed based on measurements in the UV \citep{wilson2020}.

Starting from an N$_2$-dominated atmosphere we increased the surface partial pressures of CO$_2$ from 10$^{-3}$~bar for TRAPPIST-1~e up to 10.8~bar for TRAPPIST-1~f.
The main results regarding the composition of the simulated atmospheres are listed below.

\begin{deluxetable}{p{1cm} p{3.5cm} p{3.5cm}}\label{tab:transit_bands}
    \tabletypesize{\footnotesize}
    \tablecolumns{5}
    \tablecaption{Important molecular absorption features and corresponding wavelength in $\mu$m of the simulated transmission spectra of TRAPPIST-1~e for all three scenarios and with CO$_2$-poor (10$^{-3}$, 0.01~bar) and CO$_2$-rich (0.1, 1~bar) atmospheres. In black: strong spectral features, in gray: weak spectral features.}
    \tablehead{\colhead{Scenario}  & \colhead{CO$_2$-poor (10$^{-3}$, 0.01~bar)} & \colhead{CO$_2$-rich (0.1, 1~bar)}}  
    \startdata
       ~  & O$_2$ (0.76, 1.27) & O$_2$ (0.76, 1.27) \\
       ~  & \textcolor{gray} {O$_3$ (9.6)} & O$_3$ (0.6, 9.6) \\
       Wet        &  CH$_4$ (2.3, 3.3, 7.7) &  \textcolor{gray}{CH$_4$ (2.3, 3.3, 7.7)} \\
       \& alive       &  NO$_2$ (below 0.7, \textcolor{gray}{3.45}, 6.2) & - \\
       ~   &  NO (5.3) & - \\
       ~         &  HNO$_3$ (5.85) & -  \\
       ~            &  N$_2$O (8.5) & - \\
       \hline
       Wet       &  \textcolor{gray}{O$_3$ (9.6)} & O$_3$ (9.6) \\     
       \& dead   & \textcolor{gray}{CO (2.35, 4.6)} & \textcolor{gray}{CO (2.35)} \\
       \hline
       ~             & - & \textcolor{gray}{O$_2$ (0.76, 1.27)} \\
       Dry   & - & O$_3$ (\textcolor{gray}{0.6}, 9.6) \\   
       \& dead        & CO (2.35, 4.6) & CO (2.35, 4.6) \\
        ~        &  SO$_2$ (7.35, 8.7) & SO$_2$ (7.35, 8.7) \\
    \enddata
\end{deluxetable}

\begin{itemize}
    \item The alive runs with Earth-like biogenic flux accumulate about 50\% more O$_2$ as on modern Earth due to Earth's weaker UV environment and hence weaker O$_2$ sinks. 
    
    \item For dry CO$_2$-rich atmospheres, the abiotic production of O$_2$ and O$_3$ is significant \citep[see also][]{selsis2002, segura2007, harman2015, meadows2017}, as expected due to the low FUV/NUV ratio of TRAPPIST-1 \citep{tian2014}. However, the abundances of abiotic O$_2$ and O$_3$ is one order of magnitude lower than those runs with biogenic emissions. In contrast, the wet \& dead scenario without biogenic emissions shows little abiotic O$_2$ and O$_3$ due to effective O$_2$ uptake by the ocean.
    
    \item CO can be an indirect marker of an ocean, being 100 times larger on an ocean-less world with a CO$_2$-rich atmosphere \citep[see also][]{zahnle2008, gao2015, wang2016, nava2016, meadows2017, schwieterman2019, hu2020}. 
    
    
    
    
    \item For dry scenarios the mixing ratio of O$_2$ and O$_3$ can differ by over two orders of magnitude and abundances of CO and SO$_2$ can differ by about one order of magnitude depending on the choice of the SED. For the wet scenarios the concentrations of O$_3$ in the middle atmosphere depend on the choice of the SED by a factor of $\sim$5.
    
    \item For dry scenarios the outgassed SO$_2$ leads to larger atmospheric concentrations than for the wet cases which include wet deposition.
    
\end{itemize}

We used the simulated atmospheric composition to calculate cloud-free transmission spectra of TRAPPIST-1~e for all three scenarios. Important spectral features found for the individual scenarios are listed in Table~\ref{tab:transit_bands}. 

We used the transmission spectra and the TRAPPIST-1 SED from \citet{wilson2020} to calculate the number of transits required to detect molecular features of TRAPPIST-1~e and TRAPPIST-1~f. The results are listed below.

\begin{itemize}
    \item The detection of CO$_2$ at 4.3~$\mu$m with JWST NIRSpec PRISM requires $\sim$10 transits assuming cloud-free conditions \citep[similar to findings by][]{morley2017,batalha2018, krissansen2018, wunderlich2019, lustig-yaeger2019,fauchez2019}. With the cross-correlation technique using ELT HIRES the CO$_2$ feature around 2.0~$\mu$m might be detectable by co-adding $\sim$30 transits. CO$_2$ will be easier to detect for the dry \& dead scenario due to weak absorption of H$_2$O and CH$_4$. 
    
    \item For the wet \& alive runs CH$_4$ might be detectable with ELT HIRES for the simulated cloud-free atmospheres of TRAPPIST-1~e with a surface temperature below 330~K. CH$_4$ is not detectable for any simulated case without biomass flux.
    
    \item O$_2$ is not detectable for the simulated atmospheres of TRAPPIST-1~e or TRAPPIST~1~f using the cross-correlation technique with ELT HIRES \citep[see also][]{rodler2014,serindag2019}.
    
    \item SO$_2$ indicates that a planet might not host significant surface liquid water. However, SO$_2$ is not detectable for any of the dry runs of TRAPPIST-1~e and TRAPPIST-1~f with JWST or ELT.
    
    \item CO at 2.35~$\mu$m might be detectable with JWST NIRSpec G235M for dry scenarios with weak surface deposition of CO and a CO$_2$ partial pressure above 0.01~bar. The detection of CO require about 60 transits with JWST NIRSpec PRISM and about 40 transits with ELT HIRES. The CO feature at 4.6~$\mu$m would be detectable with JWST but partially overlaps with CO$_2$ absorption. Accurate retrieval may be able to disentangle CO and CO$_2$ with JWST. 
    
\end{itemize}

We conclude that the three scenarios considered for TRAPPIST-1~e might be distinguishable for cloud-free conditions by combining $\sim$30 transit observations with JWST NIRSpec and ELT HIRES in the K-band (2.0-2.4~$\mu$m), if the CO$_2$ partial pressures on top of a 1~bar N$_2$-dominated atmosphere are above 0.01 and below 1~bar. The alive scenario, assuming Earth-like emission of CH$_4$, could be identified by the detection of CH$_4$. The non-detection of CO suggests the existence of a surface ocean. In turn, the detection of CO suggests dry surface conditions. A detection of CO$_2$ and a non-detection of CO and CH$_4$ suggests that liquid water on the surface reduces the amount of CO in the atmosphere and that biogenic emissions of CH$_4$ are weak. \\

\acknowledgments
This research was supported by DFG projects RA-714/7-1,
GO 2610/1-1, SCHR 1125/3-1 and RA 714/9-1. 
We acknowledge the support of the DFG priority programme SPP 1992 "Exploring the Diversity of Extrasolar Planets (GO 2610/2-1)". 
M. L.-P. acknowledges financial support from the State Agency for Research of the Spanish MCIU through project ESP2017-87143-R, the "Center of Excellence Severo Ochoa" award to the IAA-CSIC (SEV-2017-0709), and EC FEDER funds. PCS gratefully acknowledges support by the German Aerospace Center under DLR 50~OR~1901. J.L.G. gratefully acknowledges the support of ISSI Team 464.
We thank Micha\"{e}l Gillon for providing the SPECULOOS target list, Franklin Mills for sending cross section data of several sulphur species and Vladimir Krasnopolsky for providing chemical profiles of Mars.
The SED used is this study is based on observations made with the NASA/ESA Hubble Space Telescope, obtained from the Data Archive at the Space Telescope Science Institute, which is operated by the Association of Universities for Research in Astronomy, Inc., under NASA contract NAS 5-26555. These observations are associated with program \# 15071. Support for program \#15071 was provided by NASA through a grant from the Space Telescope Science Institute.
We thank the anonymous referee for the helpful and constructive comments.

\software{GARLIC \citep{schreier2014,schreier2018}, HITRAN 2016 \citep{gordon2017}, MPI Mainz Spectral Atlas \citep{keller2013}, ESO ETC \citep{liske2008}, NIST \citep{mallard1994}, S/N calculator for JWST \citep{wunderlich2019}, 1D Climate-Chemistry model legacy \citep[][ and others]{kasting1986,pavlov2000,segura2003,vonparis2015}}



\bibliography{references}{}
\bibliographystyle{aasjournal}



\end{document}